\documentclass[twocolumn,trackchanges]{aastex6}

\usepackage{lipsum}
\usepackage{capt-of}
\usepackage{gensymb}

\received{2017 December 19}
\revised{2018 January 25}
\accepted{2018 January 28}

\usepackage{apjfonts}
\usepackage{aas_macros, xspace} 
\usepackage{url}
\newcommand{\msun}{\mbox{$\rm M_\odot$}}
\newcommand{\Zsun}{\mbox{$Z_\odot$}}

\newcommand{\Msol}{M_{\odot}}

\renewcommand{\apjl}{ApJL}   

\usepackage{soul,xcolor}

\usepackage{hyperref}

\shorttitle{The strong gravitationally lensed system HLock01}
\shortauthors{Marques-Chaves et al.}

\begin{document}

\title{The strong gravitationally lensed $Herschel$ galaxy HLock01: Optical spectroscopy reveals a 
close galaxy merger with evidence of inflowing gas}

\author{\mbox{Rui Marques-Chaves\altaffilmark{1,2}}}
\author{\mbox{Ismael P\'{e}rez-Fournon\altaffilmark{1,2}}}
\author{\mbox{Raphael Gavazzi\altaffilmark{3}}}
\author{\mbox{Paloma I. Mart\'\i nez-Navajas\altaffilmark{1,2}}}
\author{\mbox{Dominik Riechers\altaffilmark{4}}}
\author{\mbox{Dimitra Rigopoulou\altaffilmark{5}}}
\author{\mbox{Antonio Cabrera-Lavers\altaffilmark{1,6}}}
\author{\mbox{David L. Clements\altaffilmark{7}}}
\author{\mbox{Asantha Cooray\altaffilmark{8}}}
\author{\mbox{Duncan Farrah\altaffilmark{9}}}
\author{\mbox{Rob J. Ivison\altaffilmark{10,11}}}
\author{\mbox{Camilo E. Jim\'{e}nez-\'{A}ngel\altaffilmark{1,2}}}
\author{\mbox{Hooshang Nayyeri\altaffilmark{8}}}
\author{\mbox{Seb Oliver\altaffilmark{12}}}
\author{\mbox{Alain Omont\altaffilmark{3}}}
\author{\mbox{Douglas Scott\altaffilmark{13}}}
\author{\mbox{Yiping Shu\altaffilmark{14,15}}}
\author{\mbox{Julie Wardlow\altaffilmark{16}}}

\altaffiltext{1}{Instituto de Astrof\'\i sica de Canarias, C/V\'\i a L\'actea, s/n, E-38205 San Crist\'obal de La Laguna, Tenerife, Spain}
\altaffiltext{2}{Universidad de La Laguna, Dpto. Astrof\'\i sica, E-38206 San Crist\'obal de La Laguna, Tenerife, Spain}
\altaffiltext{3}{Institut d'Astrophysique de Paris, UMR7095 CNRS \& Sorbonne Universit\'e (UPMC), F-75014 Paris, France}
\altaffiltext{4}{Astronomy Department, Cornell University, Ithaca, NY 14853, USA}
\altaffiltext{5}{Astrophysics, Department of Physics, University of Oxford, Keble Road, Oxford, OX1 3RH, UK}
\altaffiltext{6}{GRANTECAN, Cuesta de San Jos\'{e} s/n, E-38712, Bre\~{n}a Baja, La Palma, Spain}
\altaffiltext{7}{Astrophysics Group, Imperial College London, Blackett Laboratory, Prince Consort Road, London SW7 2AZ, UK}
\altaffiltext{8}{Department of Physics and Astronomy, University of California, Irvine, CA 92697, USA}
\altaffiltext{9}{Department of Physics, Virginia Tech, Blacksburg, VA 24061, USA}
\altaffiltext{10}{European Southern Observatory, Karl-Schwarzschild-Str. 2, D-85748 Garching, Germany}
\altaffiltext{11}{Institute for Astronomy, University of Edinburgh, Royal Observatory, Blackford Hill, Edinburgh EH9 3HJ, UK}
\altaffiltext{12}{Astronomy Centre, Department of Physics and Astronomy, University of Sussex, Brighton BN1 9QH, UK}
\altaffiltext{13}{Department of Physics and Astronomy, University of British Columbia, 6224 Agricultural Road, Vancouver, BC V6T 1Z1, Canada}
\altaffiltext{14}{National Astronomical Observatories, Chinese Academy of Sciences, A20 Datun Rd., Chaoyang District, Beijing 100012, China}
\altaffiltext{15}{Purple Mountain Observatory, Chinese Academy of Sciences, 2 West Beijing Road, Nanjing 210008, China}
\altaffiltext{16}{Centre for Extragalactic Astronomy, Department of Physics, Durham University, South Road, Durham DH1 3LE, UK}

\setcounter{footnote}{5}

\begin{abstract}

The submillimeter galaxy (SMG) HERMES J105751.1+573027 (hereafter HLock01) at $z = 2.9574 \pm 0.0001$ is one 
of the brightest gravitationally lensed sources discovered in the $Herschel$ Multi-tiered Extragalactic Survey. 
Apart from the high flux densities in the far-infrared, it is also extremely bright in the rest-frame ultraviolet 
(UV), with a total apparent magnitude $m_{\rm UV} \simeq 19.7$ mag.
We report here deep spectroscopic observations with the Gran Telescopio Canarias of the optically bright lensed 
images of HLock01. Our results suggest that HLock01 is a merger system composed of the $Herschel$-selected SMG 
and an optically bright Lyman break-like galaxy (LBG), separated by only 3.3 kpc in projection. 
While the SMG appears very massive~($M_{*} \simeq 5\times 10^{11}\, \rm M_{\odot}$), with a highly extinguished 
stellar component ($A_{V} \simeq 4.3 $), the LBG is a young, lower-mass ($M_{*} \simeq 1 \times 10^{10}\, 
\rm M_{\odot}$), but still luminous ($10 \times L_{\rm UV}^{*}$) satellite galaxy. 
Detailed analysis of the high signal-to-noise (S/N) rest-frame UV spectrum of the LBG shows complex kinematics of 
the gas, exhibiting both blueshifted and redshifted absorption components.
While the blueshifted component is associated with strong galactic outflows from the massive stars in the LBG, as 
is common in most star-forming galaxies, the redshifted component may be associated with gas inflow seen along a 
favorable sightline to the LBG.
We also find evidence of an extended gas reservoir around HLock01 at an impact parameter of $110$ kpc, through the 
detection of {\sc C ii} $\lambda \lambda$1334 absorption in the red wing of a bright Ly$\alpha$ emitter at $z 
\simeq 3.327$.
The data presented here highlight the power of gravitational lensing in high S/N studies to probe deeply into the 
physics of high-$z$ star forming galaxies.

\end{abstract}


\section{Introduction}

High-$z$ submillimeter  galaxies (SMGs) represent a population of the most massive and luminous galaxies in the 
early Universe. They are characterized by dust-enshrouded vigorous star formation, assembling their mass very 
rapidly over short time scales, the so-called starburst phase \citep[see][for reviews]{blain2002, casey2014}.
They are believed to be the progenitors of massive elliptical galaxies predominantly found in clusters of galaxies 
at lower redshifts \citep[e.g.,][]{daddi2009, capak2011, walter2012, dannerbauer2014, riechers2014, casey2015, 
oteo2017b}, and the study of this population is vital to understand their formation and subsequent evolution.

Despite the huge progress made over the past decades in understanding the properties of SMGs \citep[e.g.,][]{greve2005, 
magnelli2012, riechers2013, dowell2014, ivison2016, micha2017, oteo2017a}, the main mechanism that drives the 
intense star formation responsible for the high far-IR luminosities is still a matter of debate 
\citep[e.g.,][]{swinbank2008, gonzalez2011, micha2012, hayward2013, narayanan2015}. 
On one hand, almost all ultraluminous galaxies in the local Universe are interacting galaxies and mergers 
\citep[e.g.,][]{farrah2001, farrah2002, bridge2007, haan2011}. At high-$z$, a considerable number of SMGs are 
also found to be galaxy mergers \citep[e.g.,][]{capak2008, ivison2008, ivison2013, tacconi2008, fu2013, messias2014, 
rawle2014, oteo2016, marrone2018, riechers2017}, although isolated clumpy gas-rich disk galaxies can also reach 
extremely large SFRs \citep[e.g.,][]{tacconi2010, bournaud2014}.
On the other hand, some authors suggest, based on simulations, that high star-formation rates ($\rm SFRs 
\gtrapprox 1000$ $\rm \Msol$ yr$^{-1}$) in some SMGs are difficult to explain by a merger scenario alone, and propose 
that the predominant mechanism is smooth accretion of cold gas or infall of gas previously ejected via stellar 
feedback \citep[e.g.,][]{keres2005, dekel2009, narayanan2015}.

Whatever the mechanisms responsible for such extremely large SFRs are, this active phase of SMGs is a clear indication 
of intense ionizing flux from a young, massive stellar population that dominates the rest-frame ultra-violet (UV). 
It is thus important to provide detailed characterization of the physical properties of the early episodes of star 
formation in these galaxies, via spectral diagnostics in the rest-frame UV. However, such studies are limited by the 
faintness of these galaxies at short wavelengths, with the massive stars giving rise to the UV continuum being embedded 
in large quantities of dust. Considerable spectroscopic efforts have been successfully used to obtain accurate 
spectroscopic redshifts, probe for signs of active galactic nucleus (AGN) activity, and investigate properties of the 
ionized gas in SMGs, mainly by using rest-frame optical nebular emission lines \citep[e.g,][]{chapman2004, chapman2005, 
swinbank2004, swinbank2005, swinbank2006, alaghband2012, olivares2016, casey2017, danielson2017}. 
However, the typical faintness of high-$z$ galaxies and the dust obscuration of SMGs make it almost impossible with 
current facilities to obtain high signal-to-noise (S/N) spectra, a requirement for properly studying the properties 
of the young stars, and to look for signatures of outflowing/inflowing gas. 

\begin{figure*}[t!]
$\begin{array}{rl}
    \includegraphics[width=178mm]{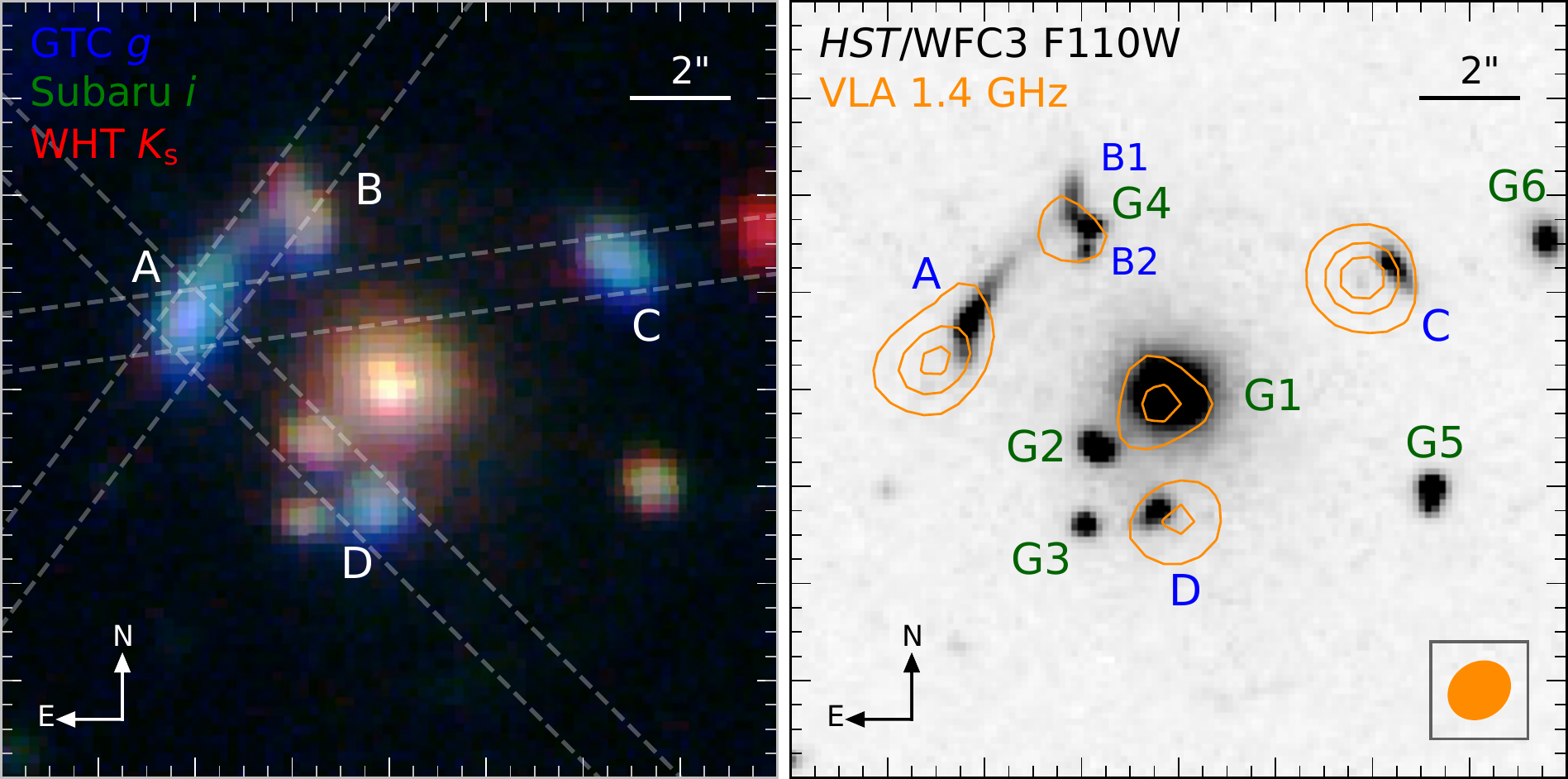}
\end{array}$
\caption{Left panel: $g$, $I$, and $K_{\rm s}$ color image of HLock01 from GTC, Subaru, and WHT, respectively. Dashed 
lines show the positions of OSIRIS long-slit spectroscopic observations, all centered on the brightest lensed image A, 
and oriented so as to encompass the other bright lensed images B, C, and D. 
Right panel: near-IR high-resolution $HST$/WFC3 F110W image with labeled multiply lensed images at $z \simeq 2.95$ (blue) 
and foreground galaxies at $z \simeq 0.65$ (green). 
G4 is massive enough to split the lensed image B into two pieces on both sides \citep[B1 and B2, see more details 
in][]{gavazzi2011}. Orange contours show VLA data at $1.4\, \rm GHz$ and its beam is shown on the bottom right. 
A spatial offset of the bright lensed images is seen between the short ($HST$ F110W, HLock01-B) and long wavelengths 
(VLA, HLock01-R). Each image is $16'' \times 16''$, centered on the brightest lensing galaxy, G1, and oriented such that 
north is up and east is to the left.
\label{fig:rgb}}
\end{figure*}

One exception to this principle is the galaxy discussed in this paper, the strong gravitationally lensed 
SMG HERMES J105751.1+573027 (hereafter HLock01), which is unusually bright in both the optical ($R \simeq 19.7$ mag) 
and in the far-IR ($S_{250\, \mu \rm m} \simeq 400$ mJy). HLock01 was identified with $Herschel$/SPIRE in the 
$Herschel$ Multi-tiered Extragalactic Survey \citep[HerMES; ][]{oliver2012}, and investigated in a series of papers 
after significant follow-up effort \citep{conley2011, gavazzi2011, riechers2011, scott2011, bussmann2013, wardlow2013}. 
Here we give a summary of the main results from those papers. 
The discovery and its lensing nature was first presented by \cite{conley2011}, based on $880\, \mu \rm m$ Submillimeter 
Array (SMA) interferometry and near-IR $K_{\rm p}$ adaptive optics (AO) observations using NIRC2 on the Keck II 
telescope, in which the $Herschel$ source was resolved into four components with a large separation of around 
$9^{\prime \prime}$  (see Figure \ref{fig:rgb}). 
Using the Plateau de Bure Interferometer (PdBI), the Combined Array for Research in Millimeter-wave Astronomy (CARMA), 
and the Green Bank Telescope (GBT), \cite{riechers2011} and \cite{scott2011} established the redshift of HLock01 
from several CO molecular emission lines as $ z_{\rm CO} = 2.9574 \pm 0.0001$. 
By studying the kinematics of the gas reservoir, \cite{riechers2011} found a resolved velocity structure in the 
CO$\, (J = 5 \rightarrow 4$) emission, similar to what is observed in gas-rich mergers, but the low spatial resolution 
did not allow a definitive conclusion.
The lens modeling was performed by \cite{gavazzi2011} using NIRC2 $K_{\rm p}$ and IRAM CO$\, (J = 5 \rightarrow 4$) imaging, 
as well as deep optical $I$-band imaging with the Subaru Telescope. 
They showed that the rest-frame UV and optical emission is magnified by a factor of $\mu = 10.9 \pm 0.7$ by a small group 
of galaxies at $z_{\rm phot} \simeq 0.6$. 
However, an offset of 2.4 kpc in the source plane was found between the stars that emit at visible/near-IR wavelengths and 
the gas distribution traced by the molecular gas. 

Later on, \cite{bussmann2013}, and \cite{wardlow2013} presented new imaging data for this system, using $Hubble$ $Space$ 
$Telescope$ ($HST$) WFC3 F110W, new $880\, \mu \rm m$ SMA with higher spatial resolution than the data presented in 
\cite{conley2011}, and Very Large Array (VLA) $1.4\,$GHz data (at $1.1^{\prime \prime}$ resolution). 
The new images show the same spatial offsets of the bright lensed images seen between the short and long wavelengths 
(see Figure \ref{fig:rgb}), noticed by \cite{conley2011} and \cite{gavazzi2011}. 
A new lens model was determined by \cite{bussmann2013} using the SMA $880\, \mu \rm m$ data, showing a large dust distribution, 
magnified by $9.2 \pm 0.4$ with an effective radius of $4$ kpc in the source plane. Its centroid matches the 
position of the gas distribution traced by the molecular gas, which we attribute to the source of the luminous far-IR 
emission, but both are offset with respect to the stars that are seen in the visible/near-IR (UV/optical in the rest-frame).
Finally, \cite{rigopoulou2018} discuss the applicability of the [O {\sc iii}]88/[N {\sc ii}]122 line ratio as a metallicity 
indicator in high redshift submillimeter luminous galaxies and found that the gas metallicity of HLock01 is 
$0.6 < Z_{\rm gas}/Z_{\odot} < 1.0$.

Due to the high dust content of HLock01, one could expect that its rest-frame UV and optical light are heavily obscured 
by dust, as in most SMGs. However, HLock01 is unusually bright in its rest-frame UV, and their colors are also consistent 
with those of $ z \sim 3$ Lyman break galaxies \citep[LBGs;][]{steidel1996} with $(G - R) = 0.5$ and $(U - G) = 1.4$.  
In this paper, we present a detailed analysis of the optically bright lensed images of HLock01, based on deep 
spectroscopic observations with the $10.4\, \rm m$ Gran Telescopio Canarias (GTC). 
Throughout the paper we adopt the name ``HLock01-B'' for the optically bright LBG-like galaxy, and ``HLock01-R'' for the 
$Herschel$-selected SMG (where ``B'' and ``R'' stand for blue and red galaxies, respectively). 
Thanks to the large collecting area of the GTC, to the lensing magnification of the source, and to the small obscuration 
towards HLock01-B, we can perform a detailed analysis of its physical properties. 

The paper is organized as follows. In Section \ref{obs}, we describe our spectroscopic and imaging observations. Our 
analysis of the rest-frame UV spectrum of HLock01-B is presented in Section \ref{uvspec}.  
The main properties of both components of HLock01, derived from SED fitting, are discussed in Section \ref{sed}. 
Finally, in Sections \ref{discussion} and \ref{conclusion}, we discus our results and summarize our main findings.
A concordance cosmology with matter and dark energy density $\Omega_{\rm m} = 0.3$, $\Omega_{\rm \Lambda} = 0.7$, 
and Hubble constant $H_{0} = 70$ km s$^{-1}$ Mpc$^{-1}$ are assumed throughout this work. All magnitudes are given 
in the AB system.

\begin{table*}[ht]
\begin{center}
\tabletypesize{\scriptsize}
\caption{OSIRIS spectroscopic observations of HLock01-B. \label{tab0}}
\begin{tabular}{c c c c c c c}
\hline \hline
\smallskip
\smallskip
Lensed images & PA & Grism & Date & Time & Seeing & Moon \\
 & ($\degree$) &  & & (sec) & (arcsec) & \\
\hline
A/B & $-39.5$ & R2500V & 2015 May 09 & $5 \times 900$ & $0.7$ & dark \\
    & $-39.5$ & R2500R & 2015 May 09 & $5 \times 900$ & $0.8$ & dark \\
A/C & $-82.2$ & R2500V & 2015 May 08 & $3 \times 900$ & $1.0$ & gray\\
    & $-82.2$ & R2500V & 2015 Jun 07 & $2 \times 900$ & $0.9$ & gray\\
    & $-82.$2 & R2500R & 2015 Apr 26 & $3 \times 900$ & $0.8$ & gray\\
    & $-82.2$ & R2500R & 2015 Jun 07 & $2 \times 900$ & $0.9$ & gray\\
A/D & 44.0 & R2500V & 2015 May 09 & $3 \times 900$ & $0.8$ & gray\\
    & 44.0 & R2500V & 2015 Jun 11 & $2 \times 900$ & $0.8$ & dark \\
    & 44.0 & R2500R & 2015 Apr 26 & $3 \times 900$ & $0.7$ & gray\\
    & 44.0 & R2500R & 2015 Jun 11 & $2 \times 900$ & $0.8$ & dark \\    
\hline 
\end{tabular}
\\
\end{center}
\textsc{      \bf{}} \\
 \\
\end{table*}

\section{Observations}\label{obs}

\subsection{GTC/OSIRIS spectroscopic and imaging observations}

Rest-frame UV spectroscopic observations were obtained with the Optical System for Imaging and low-Intermediate-Resolution 
Integrated Spectroscopy instrument (OSIRIS\footnote{\url{http://www.gtc.iac.es/instruments/osiris/}}) on the $10.4\, \rm m$ GTC.
The data used in this paper were obtained in service mode over seven different nights, between 2015 April 26 and 2015 June 21 
in dark and gray Moon conditions as part of the GTC program GTCMULTIPLE2A-15A (PI: R. Marques-Chaves).
We used the R2500V and R2500R grisms, with dispersions of 0.80 and 1.04 \AA \space px$^{-1}$, respectively. 
These two grisms provide a full spectral coverage of $4500 - 7700$ \AA, which corresponds to $1150 - 1950$ \AA \space in 
the rest-frame at $z \simeq 2.95$. 
The OSIRIS $1.2^{\prime \prime}$ wide slit was centered on the brightest lensed image of HLock01-B (image A), and oriented 
so as to encompass the other lensed images B, C, and D, at sky positions angles (PA) of $-39^{\circ}.4$, 
$-82^{\circ}.9$, and $44^{\circ}$, respectively (see Figure \ref{fig:rgb}, left panel). Given this configuration, the 
corresponding instrumental resolution for the R2500V and R2500R grisms is $\simeq 180$ km s$^{-1}$. 
In total, 15 exposures of 900 s were acquired with each grism, equally split between different PAs. 
A summary of the rest-frame UV spectroscopic observations of HLock01-B used in this work is shown in Table \ref{tab0}. 

The data were processed with standard {\sc Iraf}\footnote{\url{http://iraf.noao.edu/}} and {\sc Python} tasks. 
Each individual two-dimensional spectrum was bias-subtracted, and flat-field corrected. The wavelength calibration was 
done for every observing night using HgAr+Ne+Xe arc lamps. Finally, individual 2D spectra were background subtracted. 
The 1D spectra were then extracted and corrected for the instrumental response using observations of the standard stars 
Ross 640 and GD 153. 

We also obtained spectra of the galaxies in the group responsible for the gravitational lensing of HLock01, with two 
additional long-slit spectra to encompass $\rm G1 - G4$, and $ \rm G3 - G5$, respectively. 
For this, we used a lower spectral resolution grism, R1000R, which provides a wider spectral range ($5100 - 10000$ \AA), 
which with a $1.2^{\prime \prime}$ wide slit gives a spectral resolution of $\simeq 400$ km s$^{-1}$.
The other lensing galaxies G2, and G6 are covered by the long-slit spectra discussed before to study the lensed images 
of HLock01-B. For the galaxies G1 and G2 we detect several absorption lines (e.g., K and H of Ca~{\sc ii}~$\lambda 
\lambda$3934,3969, $\rm H_{\delta}$ $\lambda \lambda$4102, and Mgb~$\lambda \lambda$5176) as well as a prominent Balmer 
break at redshift $z_{\rm G1} = 0.6464 \pm 0.0007$, and $z_{\rm G2} = 0.6492 \pm 0.0009$, respectively. 
The spectra of G3, G4, and G5 are too noisy for a reliable measurement of their redshifts, but we marginally detect a 
jump at $6500 - 6600$ \AA, compatible with a Balmer break at $z \simeq 0.65$. 
Thus, it appears that these galaxies belong to a group at $z \simeq 0.65$, slightly larger than the previously assumed 
redshift \citep[$z_{\rm phot} = 0.6 \pm 0.04$;][]{oyaizu2008}.

Additionally, broad-band imaging with the Sloan $g'$ filter was obtained with OSIRIS on 2017 January 24, as part of the 
GTC program GTCMULTIPLE3A-16B (PI: I. P\'erez-Fournon). The total exposure time was 2160 s, spit into 12 individual 
exposures of 180 s each. Each frame was reduced individually following standard reduction procedures in {\sc Iraf}. 
The registration and combination were done using {\sc Scamp} \citep{bertin2006} and {\sc Swarp} \citep{bertin2010}. 
The seeing of the final image is $\simeq 0.8^{\prime \prime}$" (full width at half maximum, FWHM).

\subsection{WHT/LIRIS near-IR imaging}

Near-IR broad-band imaging was obtained on 2011 March 22 in the $K_{\rm s}$ filter (PI: I. P\'erez-Fournon), using the 
Long-slit Intermediate Resolution Infrared Spectrograph instrument (LIRIS) mounted at the William Herschel Telescope (WHT). 
LIRIS has a field of view of $4.27'\times 4.27'$ with a plate scale of $0.25^{\prime \prime}$ pixel$^{-1}$. The total 
integration time was 60 minutes, spit into 180 individual exposures of 20 s, adopting a random dither pattern in 15 different 
positions. The data reduction was carried out using the IAC's {\sc Iraf} 
{\sc lirisdr}\footnote{\url{http://www.iac.es/galeria/jap/lirisdr/LIRIS_DATA_REDUCTION.html}} task. 
The seeing of the final image was $0.63^{\prime \prime}$ FWHM.
The astrometric and flux calibrations were performed using 2MASS stars in the field.

\subsection{Ancillary data}

Additional data used in this work consist of a combination of shallow and deep images. Archival $U$ and $R$ 
wide-field images and catalogs from MEGACAM on the Canada-France-Hawaii Telescope (CFHT), processed and stacked 
using the MegaPipe image staking pipeline \citep{gwyn2008}, were downloaded from the Canadian Astronomy Data Centre 
(CADC\footnote{\url{http://www.cadc-ccda.hia-iha.nrc-cnrc.gc.ca/en/cfht/}}). 
Total exposure times are 4200 and 3300 s in $U$ and $R$ bands, with an average seeing
of $0.83^{\prime \prime}$ and $0.73^{\prime \prime}$ FWHM, respectively. 
HLock01 was also imaged with WFC3 on $HST$ under the snapshot program 12488 (PI: Negrello). The source was observed 
in the broad-band filter F110W with a total exposure time of 276.1 s. The $HST$ imaging was initially presented in 
\cite{bussmann2013} and \cite{wardlow2013}. 
$Spitzer$/IRAC images and catalogs in the  $3.6\, \mu$m and $4.5\, \mu$m bands from the $Spitzer$ Extragalactic 
Representative Volume Survey \citep[SERVS;][]{mauduit2012} were obtained from the IRSA 
archive\footnote{\url{http://irsa.ipac.caltech.edu/data/SPITZER/SERVS/}}. 
We also used the IRAC and MIPS imaging from the $Spitzer$ Wide-Area InfraRed Extragalactic survey data 
\citep[SWIRE;][]{lonsdale2003}.

\section{Rest-frame UV spectrum of HLock01-B}\label{uvspec}

As in other star-forming galaxies, the rest-frame UV spectrum of HLock01-B is characterized by the integrated light 
from the hot young stellar population with superimposed resonant strong absorption lines produced by the interstellar 
medium (ISM) and stellar winds. 
These spectral features can provide detailed information on dynamical, physical, and chemical properties of the 
atomic and ionized gas in the galaxy, as well as insights on the properties of the young OB stars responsible for 
the bright continuum \citep[e.g.,][]{pettini2000, pettini2002, shapley2003, jones2012, steidel2016, rigby2017a, rigby2017b}. 
Large-scale outflows of interstellar gas, resulting from the kinetic energy deposited by the star formation activity, 
are a common feature in these galaxies \citep[e.g.,][]{shapley2003, steidel2010}. 

\begin{figure*}[ht]
\centering
\includegraphics[width=180mm,scale=1]{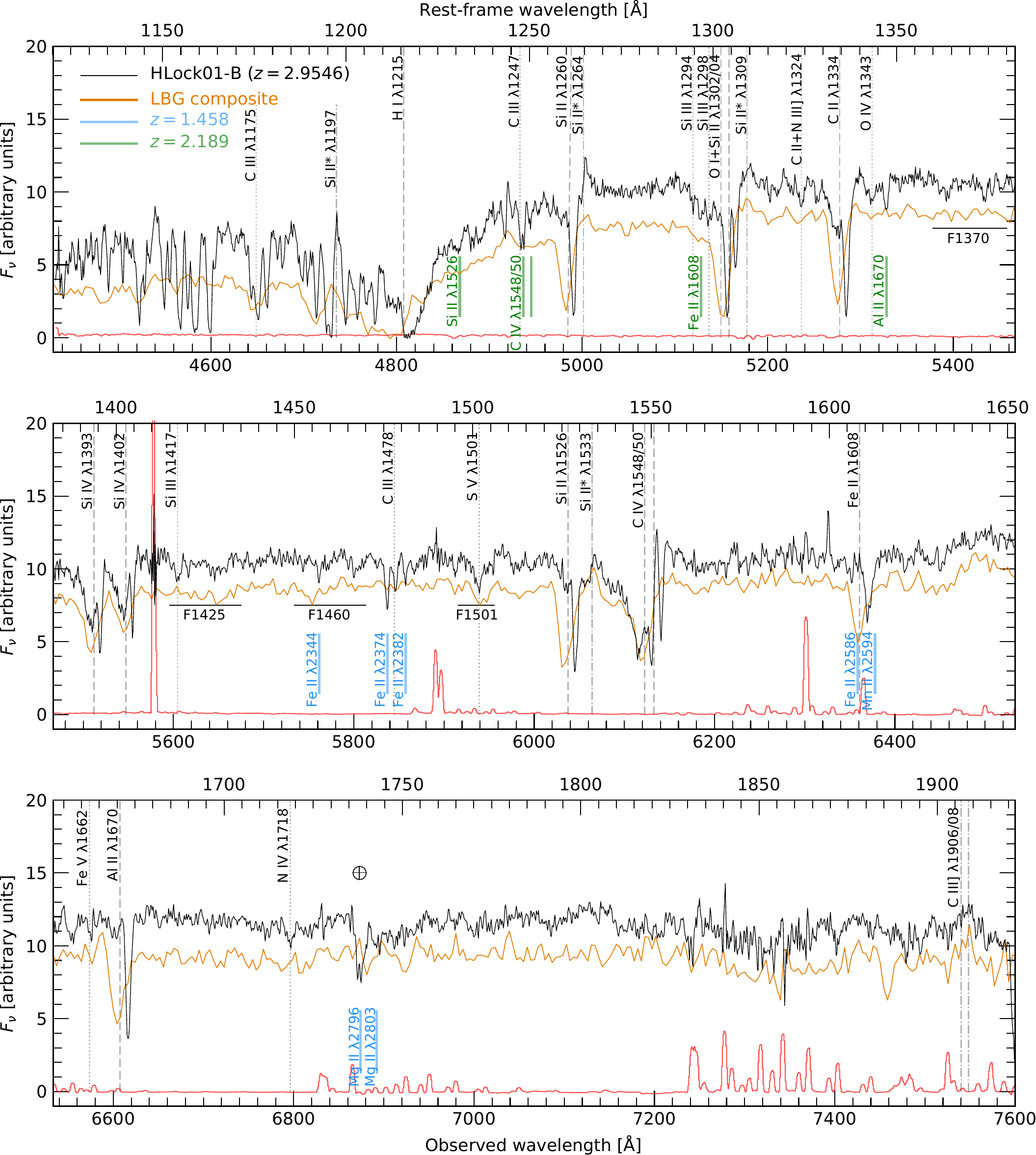}
\caption{Combined GTC/OSIRIS rest-frame UV spectrum of the lensed image A of HLock01-B. 
Vertical dotted lines identify the best defined photospheric absorption lines used to derive the systemic redshift of 
HLock01-B ($z_{\rm sys} = 2.9546 \pm 0.0004$). Strong absorption lines associated with interstellar gas and stellar 
winds are marked with vertical dashed lines. The C {\sc iii]} $\lambda \lambda$1906,1908 nebular emission doublet and 
fine-structure emission lines of Si {\sc ii} are marked with dash-dotted lines. 
For comparison, we show in orange the $z \sim 3$ LBG subset composite with the damped Ly$\alpha$ profile from 
\cite{shapley2003} at the systemic redshift of HLock01-B (downshifted for clarity).
The wavelength windows of four metallicity indices (e.g., F1370, etc.), used to derive the metallicity of young stars 
in HLock01-B, are also marked. Short vertical blue and green lines mark the positions of absorption lines of intervening 
systems at $z \simeq 1.458$ and $z \simeq 2.189$, respectively. The sky emission is also plotted in red, showing the 
locations of strong sky emission lines.
\label{fig:spec}}
\end{figure*}

Despite the differences in the S/N of the spectra of the different lensed images, there are no differences in the 
profiles of the absorption features and no evidence for velocity offsets between them, as expected. Spectra of the 
lensed images B and D show redder UV slopes ($\beta$), likely due to differential extinction from the proximity of 
their light path to the foreground galaxies. Thus, we will focus our rest-frame UV analysis on the spectrum of the 
lensed image A, which has higher S/N ($\sim 20 - 40$, depending on the wavelength range) and is less affected by 
absorption in the interstellar medium of the foreground galaxies.

The rest-frame UV spectrum of the lensed image A of HLock01-B, shown in Figure \ref{fig:spec}, is remarkably similar 
to the $ z \simeq 3$ LBG composite spectrum \citep{shapley2003}. 
It shows a damped Ly$\alpha$ absorption line, and a series of strong absorption lines associated either with stellar 
winds from massive stars (e.g., C {\sc iv} $\lambda \lambda$1548,1550), and ISM lines of several species. 
However, there are significant differences between the expected wavelengths and velocities ($\Delta v \simeq 500$ km s$^{-1}$) 
of the ISM in the $z \simeq 3 $ template (in orange in Figure \ref{fig:spec}) and in our spectrum (discussed in Section \ref{kin}). 

The OSIRIS spectrum also shows several narrow absorption lines produced by intervening systems at lower redshifts 
along the line of sight to HLock01-B. We identify at least two intervening metal systems at $z = 1.4583 \pm 0.0008$, 
and $z=2.1889 \pm 0.0007$. Some of these absorption lines may contaminate the profiles of lines of HLock01-B, and hence 
they are taken into account in our analysis (in Section \ref{kin}).

In addition, three strong Ly$\alpha$ lines at $z = 2.72$, $3.15$, and $3.27$ were serendipitously detected in 
two of our GTC long-slit spectra (see Appendix \ref{lensing}). In particular, the Ly$\alpha$ line at 
$z = 3.27$ is associated with an Ly$\alpha$ emitting galaxy at $14^{\prime \prime}$ SW from the lensing galaxy G1, 
and shows an unusual absorption line in its red emission wing consistent with C {\sc ii} $\lambda \lambda$1334 at the 
redshift of HLock01-R \citep[$ z_{\rm CO}= 2.9574$;][]{riechers2011, scott2011}. We discuss this absorption feature in 
Section \ref{reservoir}.

\subsection{Systemic redshift of HLock01-B}\label{systemic}

Stellar photospheric features are formed in the photospheres of hot stars, and although much weaker than the ISM lines, they 
can provide a measurement of the systemic redshift of the galaxy.
Within the wavelength range covered by our data, we identified several photospheric absorption features (marked with dotted 
lines in Figure \ref{fig:spec}), but some of them are blends from multiple transitions. 
Using the cleanest among these, listed in Table \ref{tab1}, we derive the mean redshift of the stars to be $ z_{\rm stars} 
= 2.9546 \pm 0.0004$. 

The nebular C {\sc III]} $\lambda \lambda$1906,1908 emission is weakly detected ($3 \sigma$) at $ z_{\rm C III]} = 2.954 
\pm 0.002$, in agreement with $z_{\rm stars}$, but the doublet is not resolved in our spectrum, and the existing data are 
too noisy for a reliable measurement of this feature.  
Therefore, throughout the paper we adopt the redshift of stellar photospheric lines as the systemic redshift of 
HLock01-B, $ z_{\rm sys} = 2.9546 \pm 0.0004$.

\begin{table}[ht]
\begin{center}
\tabletypesize{\scriptsize}
\caption{Stellar photospheric lines in HLock01-B. \label{tab1}}
\begin{tabular}{c c c c }
\hline \hline
\smallskip
\smallskip
Ion & $\lambda_{\rm lab}^{\rm a}$ (\AA)& $\lambda_{\rm obs}^{\rm b}$ (\AA)& $ z_{\rm stars}$  \\
\hline
Si~{\sc iii} & 1294.54	 & 5119.71 & $2.9548 \pm 0.0003$ \\ 
C~{\sc ii}   & 1323.93   & 5235.54$^{\rm c}$  &    $2.9543 \pm 0.0006^{\rm c}$ \\
N~{\sc iii}  & 1324.35   & 5235.54$^{\rm c}$  &    $2.9543 \pm 0.0006^{\rm c}$ \\
O~{\sc iv}   & 1343.35   & 5312.72  &    $2.9546 \pm 0.0010$ \\
Si~{\sc iii} & 1417.24   & 5604.64  &    $2.9546 \pm 0.0010$ \\
S~{\sc v}    & 1501.76   & 5939.25  &    $2.9548 \pm 0.0003$ \\
N~{\sc iv}   & 1718.55   & 6796.37  &    $2.9548 \pm 0.0004$ \\
\hline 
\end{tabular}
\\
\end{center}
\textsc{      \bf{Notes.}} \\
$^{\rm a}$ Vacuum wavelengths. \\
$^{\rm b}$ Values measured from the centroid for the individual photospheric line. \\
$^{\rm c}$ Value refers to the blended C~{\sc ii} and N~{\sc iii} photospheric lines. \\
\end{table}

\begin{figure*}[ht]
\centering
\includegraphics[width=178mm,scale=1]{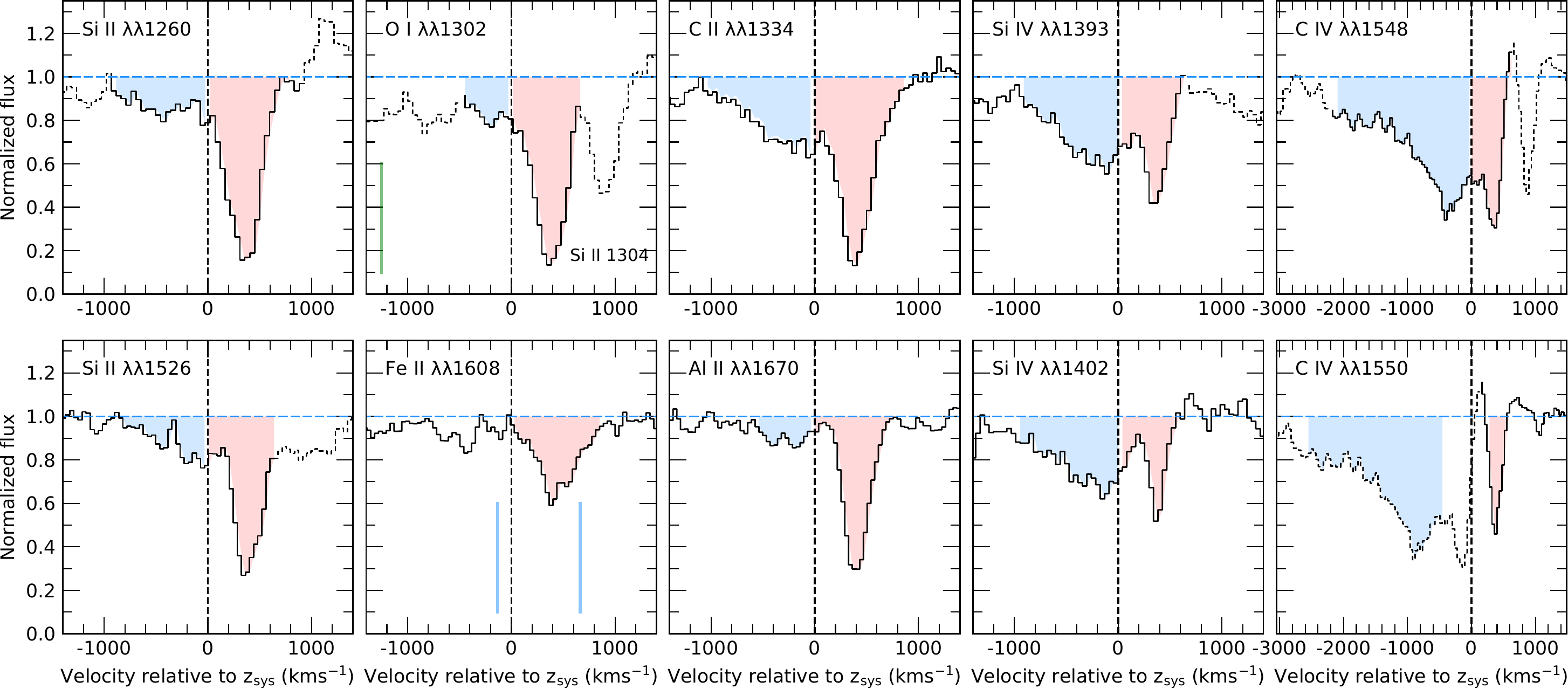}
\caption{Normalized profiles of low- (first three columns) and high-ionization (last two columns) absorption 
lines associated with the ISM and stellar winds, respectively. 
The $x$ axis is the velocity (in km s$^{-1}$) relative to the stars of HLock01-B, $z_{\rm sys} = 2.9546$, 
and the $y$ axis is the normalized flux. 
Almost all ISM and wind lines present an unusual velocity profile, with two distinct absorption components on 
either side of systemic velocity (negative and positive velocities shaded in blue and red, respectively, for 
visual purpose only). The absorption component with the maximum intensity is located at $v \simeq +370$ km s$^{-1}$ 
(shaded in red). Another broader component is also detected in high-ionization lines, like C {\sc iv} $\lambda 
\lambda$1548,1550 and Si {\sc iv} $\lambda \lambda$1393,1402, and in some ISM lines (shaded in blue).
Dotted lines indicate spectral features in HLock01 other than those to which the label in each plot refers.
Short vertical blue and green solid lines mark the expected positions of absorption lines from intervening metal 
systems at lower redshift.
\label{fig:low}}
\end{figure*}

The difference of $\Delta v = 210$ $\rm km s^{-1}$ between the systemic redshift of HLock01-B $ z_{\rm sys}= 2.9546$, 
and the redshift of HLock01-R from the molecular gas lines $ z_{\rm CO}= 2.9574$ \citep[][]{riechers2011, scott2011} 
cannot be explained by errors in redshift measurements. 
The velocity offset derived here and the complex dynamical structure of the molecular gas reservoir discussed 
in \cite{riechers2011} suggest that HLock01-B is a separate galaxy, different from the $Herschel$ SMG (HLock01-R), 
but both forming a close merger. 
Nevertheless, similar velocity offsets, interpreted as rotational velocities in some cases, have been found in a 
few massive galaxies at high-$z$ \citep[e.g.,][]{law2012, jimenez2017, toft2017}.
A more detailed discussion is presented in Section \ref{merger}. \\

\subsection{Kinematics of the ISM}\label{kin}

Within our spectral range, we identify 11 strong absorption features, including low-ionization lines (Si {\sc ii} 
$\lambda \lambda$1260, O {\sc i} $\lambda \lambda$1302, Si {\sc ii} $\lambda \lambda$1304, C {\sc ii} $\lambda 
\lambda$1334, Si {\sc ii} $\lambda \lambda$1526, Fe {\sc ii} $\lambda \lambda$1608, and Al {\sc ii} $\lambda 
\lambda$1670), and high-ionization lines associated with a hot gas phase (Si {\sc iv} $\lambda \lambda$1393,1402, 
and C {\sc iv} $\lambda \lambda$1548,1550).
In low-ionization lines, the interstellar component usually dominates over the stellar contribution, and thus they 
are useful for studying the kinematics of the ISM \citep{shapley2003, steidel2010}. High-ionization lines are associated 
with strong winds from young stars, and predominantly trace gas at higher temperatures ($T \geq 10^{4}$ K).

For the kinematic analysis of the ISM, we firstly normalized the GTC/OSIRIS spectrum of HLock01-B using the 
pseudo-continuum windows that are free of absorption and emission features identified by \cite{rix2004}.

Figure \ref{fig:low} shows the normalized profiles of the strongest absorption lines seen in our spectrum. 
We note that all ISM lines present an unusual velocity profile with the maximum optical depth located at a mean 
$v = (+370 \pm 30)$ km s$^{-1}$ relative to the stars of HLock01-B, or $v = (+170 \pm 30)$ km s$^{-1}$ relative 
to the $Herschel$ SMG at $z_{\rm CO} = 2.9574$. 
This can be understood as gas apparently moving towards the young stars, since all the interstellar lines are seen 
against the UV stellar continuum. 
This absorption is strong in the low-ionization lines (the first three columns in Figure \ref{fig:low}), likely 
with saturated profiles~\footnote{We test if some of these lines are saturated, by considering the linear part of 
the curve of growth. In this case, the ratios of the rest-frame equivalent widths ($\rm EW_{0}$) of different 
transitions of a given ion can be related through their oscillator strengths. For example, for the Si {\sc ii} 
lines in the unsaturated case we would expect $\rm EW_{0} (1260)$ / $\rm EW_{0} (1526) \simeq 5$, $\rm EW_{0} 
(1260)$ / $\rm EW_{0} (1304) \simeq 10$, and $\rm EW_{0} (1304)$ / $\rm EW_{0} (1526) \simeq 0.5$. 
Our spectrum shows ratios of $\simeq 1.0, 2.5$, and $0.4$, respectively, suggesting that at least Si {\sc ii} 
$\lambda \lambda 1260$ may be saturated.}, but it is also present, although notably weaker in high-ionization 
ones, like C {\sc iv} and Si {\sc iv} (the last two columns in Figure \ref{fig:low}).  
The spectrum of HLock01-B also shows a secondary, but broader absorption component centered at a mean $v = 
(-220 \pm 60)$ km s$^{-1}$ relatively to its systemic redshift, which is a characteristic of large-scale outflows 
of material in HLock01-B. This blueshifted component is stronger in high-ionization lines than in the low-ionization 
ones (it is detected in C {\sc ii} $\lambda \lambda$1334,  Si {\sc ii} $\lambda \lambda$1260, and Si {\sc ii} $\lambda 
\lambda$1526, but is not clear in O {\sc I} $\lambda \lambda$1302, Fe {\sc ii} $\lambda \lambda$1608 or Al {\sc ii}$\lambda 
\lambda$1670), suggesting that the outflowing gas is mostly ionized, or the neutral gas has a lower covering factor than the ionized gas.

The absorption profiles resulting from these two components extend over a velocity range $\Delta v 
\simeq 1700$ km s$^{-1}$, from $\sim -1000$ to $\sim +700$ km s$^{-1}$, much larger than in other high-$z$ lensed 
LBGs \citep[][]{pettini2000, pettini2002, cabanac2008, quider2009, quider2010, des2010}. 
The C {\sc iv} doublet is even broader than the ISM lines, with $\Delta v \gtrsim 3000$ km s$^{-1}$, 
indicative of a strong contribution from  winds due to radiation pressure of the most massive, and luminous stars of HLock01-B. 
The velocity profile of the C~{\sc iv} doublet shows a strong P-Cygni profile, with the red-emission wing being 
attenuated by the two narrow, redshifted ($v \simeq +370$ km s$^{-1}$) interstellar absorption components of C~{\sc iv}.

\begin{figure}[ht]
\centering
\includegraphics[width=85mm,scale=1]{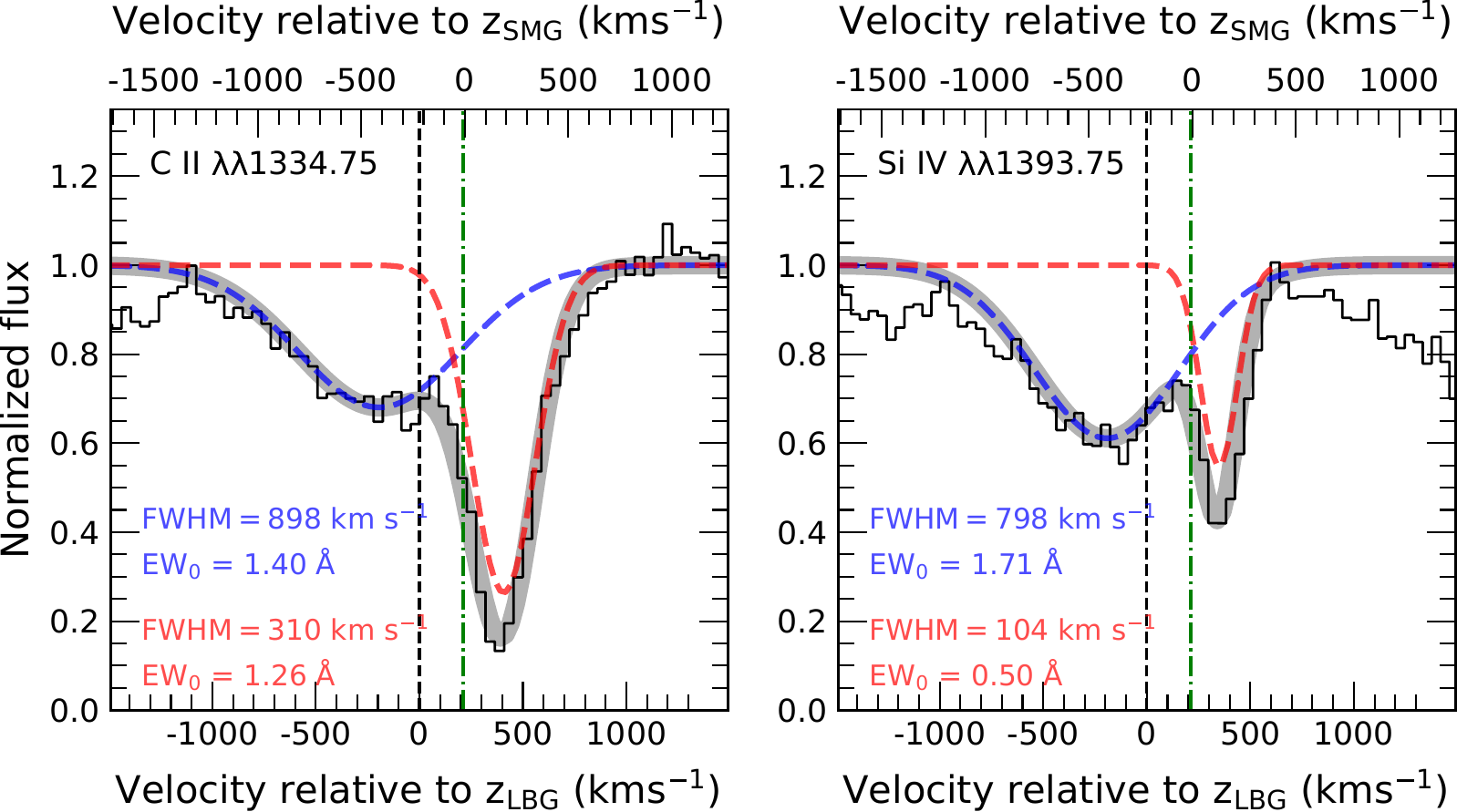}
\caption{
Example of the double-Gaussian fit of the low- (C {\sc ii} $\lambda \lambda$1334) and high-ionization lines (Si~{\sc iv} 
$\lambda \lambda$1393). Lower and upper $x$ axes are the velocity (in km s$^{-1}$) relative to the stars of the LBG 
(HLock01-B) and to the molecular gas of the SMG (HLock01-R), respectively. The $y$ axis is the normalized flux. 
The redshifted (red color) and blueshifted (blue color) components represent outflowing and inflowing gas along the 
line of sight of HLock01-B, respectively. The sum of the two components is also shown in gray. 
Black dashed and green dot-dashed vertical lines mark the zero velocity position with respect to $z_{\rm sys} = 2.9546$ 
(HLock01-B) and $z_{\rm CO} = 2.9574$ (HLock01-R), respectively.\label{fig:gauss}}
\end{figure}

\begin{table*}[htbp]
\begin{center}
\caption{\label{tb:tb1} Strong Interstellar Absorption Features in the Spectrum of HLock01-B.}
\begin{tabular}{c c c c c c c c c c c}
\hline \hline
Ion & $\lambda_{\rm lab}$  & $\Delta v$  & $\rm EW_{0}^{\rm total}$  & \multicolumn{3}{c}{Blue (Outflowing)} & & 
\multicolumn{3}{c}{Red (Inflowing)} \\
\cline{5-7} \cline{9-11} & & & & $z_{\rm b}$ & $\rm EW_{0}$  & FWHM  & & 
$z_{\rm r}$ & $\rm EW_{0}$  & FWHM  \\ 
 & (\AA) & (km s$^{-1}$) & (\AA) & & (\AA) & (km s$^{-1}$) & & & (\AA) &  (km s$^{-1}$) \\
(1) & (2) & (3) & (4) & (5) & (6) & (7) & & (8) & (9) & (10) \\ 
\hline 
Si~{\sc ii}	& 1260.42	&	$-950 \pm 750$ & $1.94 \pm 0.23$  & $2.9503 $ & $0.69 \pm 0.20$ & 900 & & $2.9596 $ & $1.32 \pm 0.26$ & 309 \\
O~{\sc i} 	& 1302.17	& 	$-950 \pm 750$ & $2.52 \pm 0.21$  & ---  & --- & --- & & $2.9598 $ & $1.61 \pm 0.24$ &  373 \\
Si~{\sc ii} & 1304.37	& 	$-950 \pm 750$ & $2.52 \pm 0.21$  &--- & --- & --- & & $2.9597 $ & $0.76 \pm 0.12$ & 244 \\
C~{\sc ii} 	& 1334.53	& 	$-1000 \pm 900$ & $2.59 \pm 0.26$  & $2.9520 $ & $1.40 \pm 0.20$ & $898 $ & & $2.9599 $ & $1.26 \pm 0.12$ & 310 \\
Si~{\sc ii} & 1526.71	& 	$-900 \pm 700$ & $1.82 \pm 0.27$  & $2.9531 $ & $0.76 \pm 0.15$ & 740 & & $2.9596 $ & $1.09 \pm 0.16$ & 230 \\
Fe~{\sc ii} & 1608.45	& 	$-800 \pm 950$ & $1.27 \pm 0.36^{\rm a}$  & --- & --- & --- & & 2.9607$^{\rm a}$ &  $1.07 \pm 0.21^{\rm a}$ &  468$^{\rm a}$     \\
Al~{\sc ii} & 1670.79	& 	$-800 \pm 750$ & $1.58 \pm 0.44$  & 2.9510 & $0.43 \pm 0.20$ & 554 & & 2.9599 & $1.16 \pm 0.13$ & 198 \\
Si~{\sc iv}	& 1393.76	& 	$-1000 \pm 650$ & $2.25 \pm 0.25$  & $2.9520 $ & $1.71 \pm 0.19$ & 798 & & $2.9592 $ & $0.50 \pm 0.10$ & 104 \\
Si~{\sc iv}	& 1402.77   & 	$-1000 \pm 650$ & $1.64 \pm 0.27$  & $2.9514 $ & $1.45 \pm 0.15$ & 859 & & $2.9599 $ &  $0.29 \pm 0.08$ & <180   \\
C~{\sc iv}  & 1548.20   & 	$-2600 \pm 1000$  & $5.76 \pm 0.40$  & --- & --- & --- & & $2.9593 $ & $1.07 \pm 0.40$ & <180 \\
C~{\sc iv}  & 1550.78   & 	$-2600 \pm 1000$  & $5.76 \pm 0.40$  & --- & --- & --- & & $2.9596 $ & $0.54 \pm 0.20$ & <180 \\
\hline 
\label{table:lines}
\end{tabular}
\end{center}
\textsc{      \bf{Notes.}} --- Columns are as follows: (1) and (2) ion and the corresponding vacuum wavelength; (3) velocity 
range for the measurements of the total rest-frame equivalent width (blueshifted and redshifted components); (4) total 
rest-frame equivalent width and $1 \sigma$ error; 
(5), (6) and (7) redshift, rest-frame equivalent width, and full width half maximum ($\rm FWHM = 2 \sqrt{2 \rm ln 2} \sigma$) 
of the blueshifted absorption component from the Gaussian fit; 
(8), (9), and (10) are the same as (5), (6), and (7), but in this case applied to the redshifted absorption component from 
the Gaussian fit. \\
$^{\rm a}$ The profile of Fe {\sc ii} $\lambda \lambda$1608 is affected by other absorption lines from an intervening system at 
$z = 1.4583 \pm 0.0005$. 
\\ 
\end{table*}

In order to understand the blueshifed and redshifted ISM absorption components, we simultaneously fit two Gaussians to the 
low- and high-ionization lines. Figure \ref{fig:gauss} shows an example of our fit to the low-ionization C {\sc ii} $\lambda \lambda$1334, 
and high-ionization Si {\sc iv} $\lambda \lambda$1393 lines, and Table \ref{table:lines} summarizes the results for all strong absorption 
lines, except the ones that are affected by the proximity of other lines ({\sc O i} $\lambda \lambda$1302, Si {\sc ii} $\lambda \lambda$1304, 
and Fe {\sc ii} $\lambda \lambda$1608). In C {\sc iv} $\lambda \lambda$1548,1550 we only fit the redshifted component, 
as the blueshifted one is affected by strong winds.

The fitted blueshifted component (blue dashed lines in Figure \ref{fig:gauss}) has its peak located at $z_{\rm blue} = 2.9519 \pm 0.0009$ 
(or $v_{\rm blue} \simeq -220$ km s$^{-1}$ relative to the systemic redshift of HLock01-B), and shows a broad profile ($\rm FWHM \simeq 900$ 
km s$^{-1}$, after accounting for the instrumental broadening) extended over a velocity range from about $-1000$ 
to $+600$~km~s$^{-1}$. Despite the fact that a potentially large contribution from stellar winds may be 
present in the high-ionization lines, for the low-ionization ones, this effect is negligible. 
Thus, the broadness of the blueshifted component in the low-ionization line C {\sc ii} (as well as in others like 
Si {\sc ii} $\lambda \lambda$1260, Si {\sc ii} $\lambda \lambda$1526, see Figure \ref{fig:low}) suggests highly turbulent 
kinematics of the outflowing gas. 
As a comparison, the velocity profile of the low-ionization interstellar lines in the $ z \sim 3$ LBGs composite spectrum of 
\cite{shapley2003} shows an average $\rm FWHM = (560 \pm 150)$~km~s$^{-1}$. 
Despite the broadness of the outflowing low-ionization lines in HLock01-B, they show smaller rest-frame equivalent widths 
(by a factor of two) than in typical LBGs.

The fitted redshifted component (red dashed line in Figure \ref{fig:gauss}) has its peak located at $z_{\rm red} = 2.9596 
\pm 0.0003$, and shows a narrow profile (FWHM $=310$ and $104$ km s$^{-1}$, for C {\sc ii} and Si {\sc iv}, respectively). 
If dynamically related to HLock01, its positive velocity relatively to both HLock01-B ($v \simeq +370$ km s$^{-1}$) and 
HLock01-R ($v \simeq +170$ km s$^{-1}$) is indicative of gas moving towards the system. 
This narrower component is seen in all absorption lines, but appears almost unresolved in the high-ionization ones, 
suggesting that these lines are dominated by the interstellar component. Moreover, the differences in the FWHM between 
the redshifted component in low- and high-ionization lines (see Table \ref{table:lines}) also suggests that cold and warm 
gas may have different kinematics and origins. 
A more clear picture of the origin of this redshifted component of the ISM may come from deeper and higher spectral resolution 
observations. In particular, fully resolved ISM components of HLock01-B with unsaturated profiles may be used to 
derive chemical abundances, using the apparent optical depth method, as has been done in other studies of strongly lensed 
star-forming galaxies \citep[e.g.,][]{pettini2002, quider2009, des2010}.

\subsection{Stellar metallicity and age of HLock01-B}\label{met}

\cite{leitherer2001}, \cite{rix2004}, and later on \cite{sommariva2012}, showed that several blends of 
stellar UV photospheric absorption lines can be used to trace the metallicity of young stars, by measuring equivalent 
widths of these blends in specific wavelength windows. These metallicity indicators have been successfully applied in several 
works \citep[e.g.,][]{quider2009, quider2010, patricio2016}, using high S/N spectra due to the faintness of these absorption lines. 
They are defined as the F1370, F1425, F1460, F1501, and F1978 indices, and their wavelength windows are shown in 
Figure~\ref{fig:spec} (except for the latter one which is not covered by our data).

We measured the equivalent widths of the features in each of these spectral windows using our normalized spectrum. We then 
applied the calibrations of \cite{sommariva2012} to obtain the corresponding metallicity. Table \ref{tab_index} summarize our 
measurements. All indices agree in a sub-solar metallicity, and we use the mean value to derive the metallicity of the UV stars 
in HLock01-B as $Z_{\rm stars} = (0.4 \pm 0.1) \Zsun$.

\begin{table}[ht]
\begin{center}
\tabletypesize{\scriptsize}
\caption{Metallicity estimates following \cite{sommariva2012}. \label{tab_index}}
\begin{tabular}{c c c c }
\hline \hline
\smallskip
\smallskip
Index & Range (\AA) & $\rm EW$ (\AA) & $Z / \Zsun$ \\
\hline
F1370 & 1360 -- 1380 & $1.4 \pm 0.2$ & 0.35 \\
F1425 & 1415 -- 1435 & $0.9 \pm 0.1$ & 0.28 \\
F1460 & 1450 -- 1470 & $1.0 \pm 0.1$ & 0.39 \\
F1501 & 1496 -- 1506 & $0.6 \pm 0.1$ & 0.59 \\
\hline 
\end{tabular}
\end{center}
\end{table}

The strength of any P-Cygni features formed in the expanding winds of the most massive stars is also sensitive to the 
metallicity, along with the age and the initial mass function (IMF) of the stellar population. 
In order to study the P-Cygni stellar wind features in HLock01-B we use model spectra computed with the spectral 
synthesis code {\sc Starburst99} \citep{leitherer1999, leitherer2001} and perform $\chi ^{2}$ minimization to our 
normalized data. In generating the {\sc Starburst99} spectra, we assumed both continuous and instantaneous star-formation 
scenarios. We adopt a Salpeter slope for the IMF between 1 and 100~$\rm \msun$, and ages ranging from 1 to 100~Myr. 
{\sc Starburst99} models were generated using libraries of empirical UV spectra from Large and Small Magallenic Cloud stars (LMC/SMC, 
which correspond to a metallicity $Z_{\rm MC} \simeq 0.4 \Zsun$) and Galactic stars ($\simeq 1 \Zsun$). 

\begin{figure*}[ht]
\centering
\includegraphics[width=170mm,scale=1]{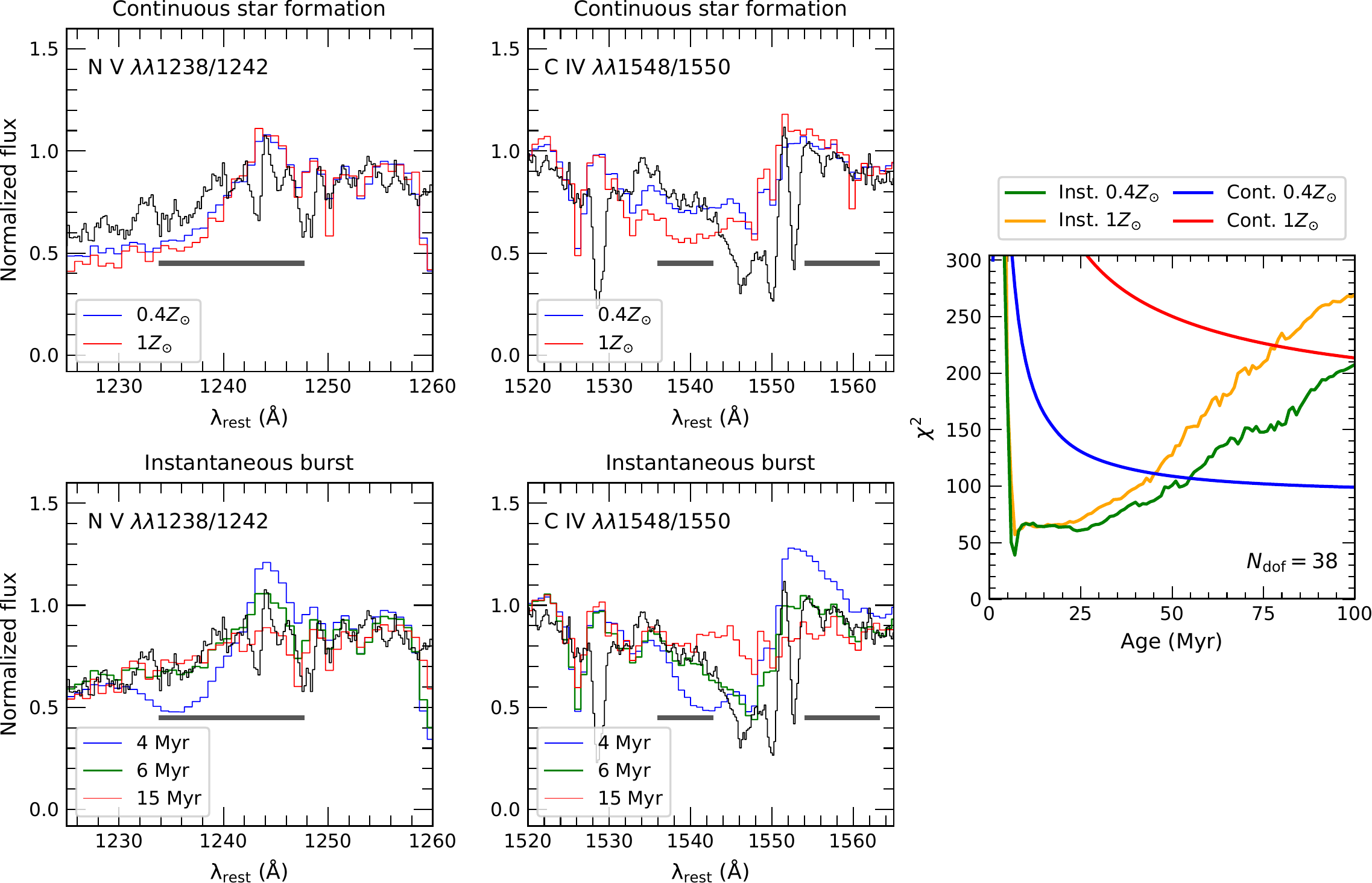}
\caption{Comparison of the normalized spectrum of HLock01-B (black solid lines) with empirical models from {\sc Starburst99} 
for high-ionization lines associated with stellar winds: N~{\sc v}~$\lambda \lambda$1238,1242 (left panels); 
%Si {\sc iv} $\lambda \lambda$1393,1402 (middle panels), 
and C~{\sc iv}~$\lambda \lambda$1548,1550 (middle panels). Upper panels show 
models for continuous star formation with a stellar population age $t = 100$ Myr, and for metallicities of $0.4 \Zsun$ 
(LMC/SMG stars, blue solid lines) and $1 \Zsun$ (Galactic stars, red solid lines). 
Lower panels show the comparison of the spectrum of HLock01-B with models for an instantaneous burst with $Z = 0.4 \Zsun$ and ages of 
$4$ (blue), $6$ (green), and $15$ (red) Myr. The wavelength windows used to perform the $\chi^{2}$ minimization are marked in gray. 
Note that we exclude in this fit the region of strong interstellar absorption in C {\sc iv} (between $1543$ and $1553$). 
The right panel shows the $\chi^{2}$ ($N_{\rm dof} = 38$) as a function of the age of the stellar population for continuous and instantaneous 
star formation models, and metallicities of $0.4 \Zsun$ and $1 \Zsun$, as indicated. 
A good fit is seen for models using LMC/SMG stars ($0.4 \Zsun$) with around $6$ Myr old burst.
\label{fig:s99}}
\end{figure*}

Figure \ref{fig:s99} shows two high-ionization lines associated with stellar winds used in this fit: N~{\sc v}~$\lambda 
\lambda$1238,1242 (left panels); and C~{\sc iv}~$\lambda \lambda$1548,1550 (middle panels). 
{\sc Starburst99} models are plotted for continuous (upper panels) and instantaneous (lower panels) star-formation scenarios. 
We excluded in the fit the region encompassing the interstellar absorption in C~{\sc iv}, between $1543$ and $1553$ \AA, which 
is not associated with the stellar P-Cygni profile.

The $\chi^{2}$ ($N_{\rm dof} = 38$) as a function of the age of the stellar population for all models is shown in the right 
panel of Figure \ref{fig:s99}. The model using LMC/SMG stars ($0.4 \Zsun$) with a $\sim 6$ Myr old burst matches the data better 
($\chi^{2} / N_{\rm dof} = 39/38$), reproducing quite well both the line profiles of N~{\sc v} and C~{\sc iv}, in particular the 
regions of the blueshifted absorption wind component of the {\sc C iv} P-Cygni profile and its red-emission wing (regions of 
$1536$ to $1542$ and $1554$ to $1563$ \AA, respectively).
Bursts over sightly longer periods ($\lesssim 25$ Myr) can also recover the stellar blueshifted absorption of 
{\sc C iv}, but fail to reproduce the red emission wing of HLock01-B (see lower panels of Figure \ref{fig:s99} for an example 
of a $15$ Myr old burst). However, our results should be treated with care, given the assumptions made on the slope and 
upper end cut-off of the IMF. Moreover, the P-Cygni profile of N~{\sc v}~$\lambda \lambda$1238,1242 could also be affected 
by the red wing of the damped Ly$\alpha$ absorption line (see Section \ref{damp}).

All methods used in this work to derive the metallicity of the young stars in HLock01-B point to a $Z_{\rm stars} 
\simeq 0.4 Z_{\odot}$ value. Our metallicity measurements are slightly different from the measurement of gas metallicity in 
HLock01-R ($0.6 < Z_{\rm gas}/Z_{\odot} < 1.0$) by \cite{rigopoulou2018}, using the [O {\sc iii}]88/[N {\sc ii}]122 line ratio 
as a metallicity indicator. This provides additional evidence that the bright $Herschel$-SMG (HLock01-R) and the bright 
LBG-like galaxy (HLock01-B) are likely different galaxies with distinct enrichment histories.

\subsection{The damped Ly$\alpha$ profile}\label{damp}

\begin{figure}[ht]
\centering
\includegraphics[width=85mm,scale=1]{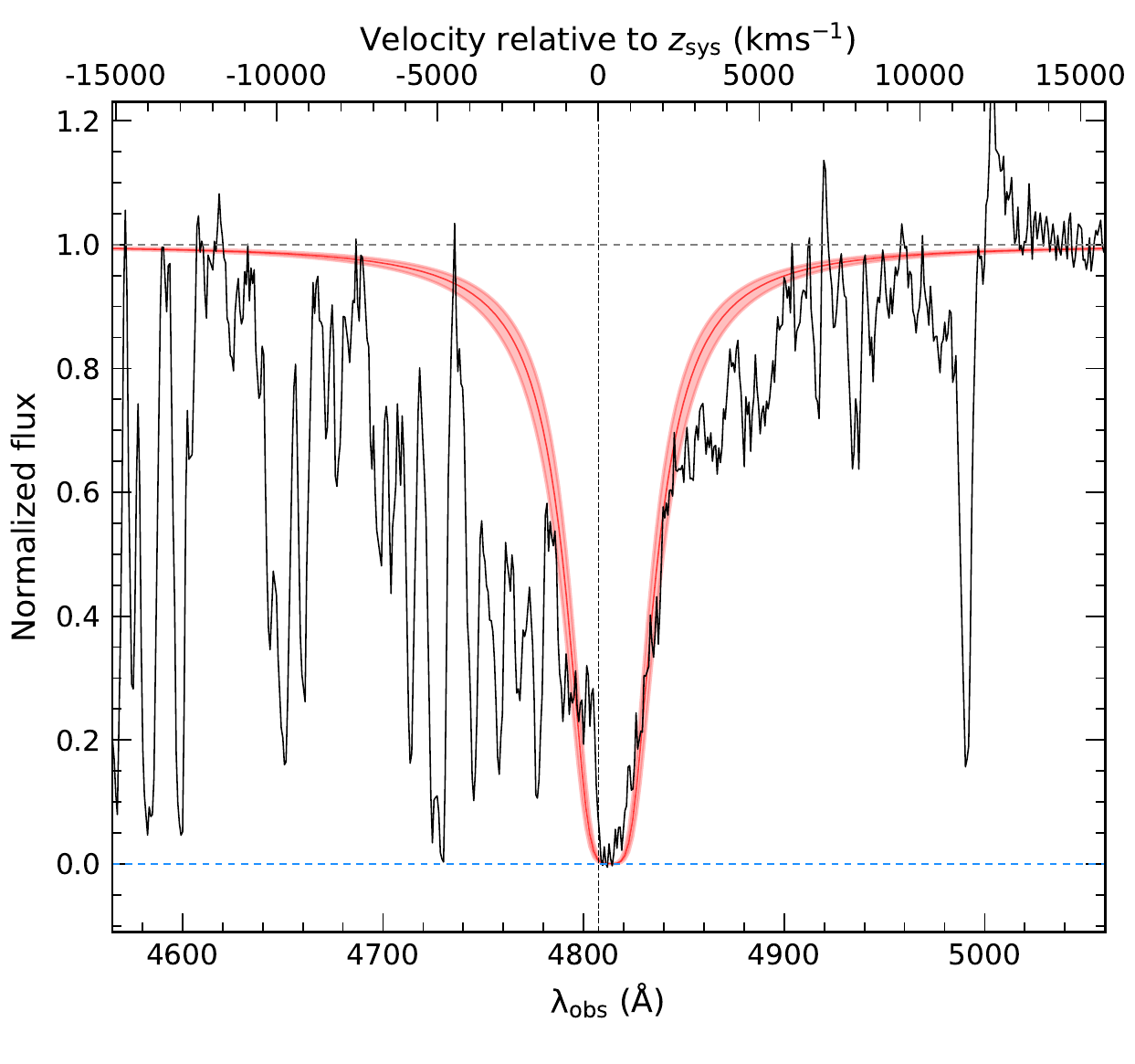}
\caption{Portion of the spectrum of HLock01-B encompassing the region of the Ly$\alpha$ absorption line. The $x$ axis is the 
observed wavelength (in \AA) and the $y$ axis is the normalized intensity. Overlaid is the best Voigt profile fit (red solid line) 
for $ N $(H {\sc i}) $=( 5.83 \pm 1.24) \times 10^{20}$ cm$^{-2}$ centered at $z_{\rm Ly \alpha} = 2.9594$, as well as the $1 \sigma$ 
error (red filled area).
\label{fig:lya}}
\end{figure}

The Ly$\alpha$ line in the spectrum of HLock01-B shows a strong damped Ly$\alpha$ profile, with the minimum lying in the range 
from $\simeq 4809$ to $4815$ \AA \space (or $v \simeq +100$ to $+500$ km s$^{-1}$ relative to $z_{\rm sys}$, see Figure \ref{fig:lya}). 
We used the software {\sc PyAstronomy}\footnote{\url{https://github.com/sczesla/PyAstronomy}} to generate theoretical 
Voigt profiles and perform $\chi ^{2}$ minimization to the Ly$\alpha$ profile.
However, we noted differences in the absorption profile in the blue and red damping wings, with the absorption being more 
pronounced in the latter. The blue wing is more noisy and less constrained than the red one, likely due to additional absorption 
from the intergalactic medium in the line of sight towards HLock01-B. The red wing is well fitted with a neutral hydrogen column 
density $ N $(H {\sc i}) $=( 5.83 \pm 1.24) \times 10^{20}$ cm$^{-2}$ centered at $\simeq 4813$\AA $ $ or $v \simeq +370$ 
km s$^{-1}$ with respect to $z_{\rm sys}$. 
Therefore, we interpret that most of the damped absorption is due to the redshifted component of the ISM seen in the spectrum of 
HLock01-B, which consists primarily of neutral gas, and presents a large optical depth. Our derived column density of H {\sc i} 
is in the range of typical values measured in other lensed galaxies \citep[e.g.,][]{pettini2000, cabanac2008, des2010}.

\subsection{Weak emission lines}

As discussed in Section \ref{systemic}, the nebular C~{\sc iii]}~$\lambda \lambda$1906,1908 emission is barely detected 
in our OSIRIS spectrum, and the doublet is not resolved. Despite the low significance of the detection ($3 \sigma$) we 
fit a Gaussian to the unresolved C {\sc iii]} doublet. We derive $z_{\rm C III]} = 2.954 \pm 0.002$ and $\rm EW_{0}^{\rm 
C III]} = (1.0 \pm 0.3)$ \AA. 
Other semiforbidden transitions often detected in the spectra of star-forming galaxies are the O~{\sc iii]}~$\lambda 
\lambda$1661,1666 lines, but these are not detected in our spectrum, despite the high continuum S/N in this spectral 
region (around $25$). The absence of these nebular lines in HLock01-B may be due to their faintness or to contamination 
with the overlapping blueshifted component of the Al~{\sc ii} $\lambda \lambda$1670 interstellar line.
In addition to C~{\sc iii]}, we also detect emission features from the excited fine-structure transitions
Si~{\sc ii*}~$\lambda \lambda$1264, 1309, and 1533. Their profiles appear slightly asymmetric (see Figure \ref{fig:spec}),
with the centroids redshifted with respect to $z_{\rm sys}$ by a mean of $\simeq +120 \pm 50$~km~s$^{-1}$. 
This velocity offset may be due to the neighboring red component of resonance absorption features 
(Si~{\sc ii}~$\lambda \lambda$1260, O~{\sc i}~+~Si~{\sc ii}~$\lambda \lambda$1303, and Si~{\sc ii}~$\lambda \lambda$1526), which 
attenuate the blue edges of the fine-structure emission profiles. 
We measure rest-frame equivalent widths of $0.32\pm0.06$, $0.22\pm0.09$, and $0.26\pm0.14$ \AA \space for 
Si~{\sc ii*}~$\lambda \lambda$1264, 1309, and 1533, respectively. \\

\section{Physical Properties from the spectral energy distribution}\label{sed}

\cite{conley2011} analyzed the SED of HLock01, considering only a single lensed background source. 
They simultaneously fitted the emission in the optical/near-IR with longer wavelength data (far-IR and submm), but the 
fit did not explain the IRAC fluxes and overestimated the $K_{\rm p}$ and MIPS 70$\mu$m flux densities by a factor of 2 or more.
We now know that HLock01 is composed of two different, spatially and spectrally offset sources, and thus the energy balance method 
(between dust-absorbed stellar continuum and the reprocessed dust emission in the far-IR), which was previously invoked, 
cannot be applied to the integrated photometry by considering a single source. 

From a revised photometric analysis of both background sources, presented in Appendix \ref{bb}, we show that their SEDs 
are well defined at short (HLock01-B) and long wavelengths (HLock01-R), where these two components dominate, respectively. 
HLock01-B is very bright in the rest-frame UV and optical, but faint (or undetected) in the current VLA and SMA data.  
On the other hand, HLock01-R shows a very red and obscured counterpart in the rest-frame UV and optical, but is very bright 
in the submm. However, to deblend the emission from the two components in the IRAC bands is challenging, given the limitation 
of the low spatial resolution. 
The centroids of the bright lensed images A and C seen in the IRAC 3.6 and 4.5 $\mu$m bands are slightly offset 
($\simeq 0.7^{\prime \prime}$) with respect to the bright counterparts in the optical and submm, suggesting a contribution 
of both to the total flux density in IRAC bands. 
However, the small rest-frame UV spectral slope of HLock01-B, $\beta = -1.9 \pm 0.1$ (measured from the observed $R$ and $I$ 
bands, and assuming a simple power law $F_{\lambda} = \lambda^{\beta}$), and the low Balmer/$4000$ \AA \space break color 
($\rm F110W -$ $K_{\rm s} = 0.15$ mag), suggest that the contribution of the LBG in the mid-IR is modest compared with the 
emission from the SMG.
Also, typical LBGs are faint in the mid-IR \citep[$S_{24 \mu \rm m} \simeq 20-30$ $\mu \rm Jy$;][]{magdis2010, reddy2012}, 
even those showing a redder UV $\beta$ slope \citep{reddy2006, coppin2007, siana2008, siana2009, reddy2010, magdis2017}, 
which are on average more massive and show larger infrared luminosities. 

Additionally, in strong gravitational lensing the finite extend of one or multiple background sources can lead to significant 
differential magnification, and their intrinsic properties, derived from photometric or spectroscopic diagnostics, 
can be incorrect if this effect is not taken into account \citep[e.g.,][]{hezaveh2012, serjeant2012}. 
Of particular importance in treating differential magnification are the cases with multiple background sources with 
significantly different SEDs and positions in the source plane \citep[e.g., the $z\sim 2.9$ gravitational lensed system 
analyzed in][]{mackenzie2014}. To check this effect in HLock01, we use the high S/N and high spatial resolution $HST$/F110W 
imaging data to update the lens model already described in \cite{gavazzi2011}. The procedure is detailed in Appendix~\ref{lens}. 
Figure~\ref{fig:offset} shows the mean positions and ellipses characterizing the galaxy shapes in the source plane for all wavebands. 
Stellar emission in the optical/near-IR (LBG) is coincident and slightly offset by $0\farcs42 \pm 0\farcs07$ in the source 
plane from the mutually coincident VLA, CO, and dust emission (SMG). 
We find magnification factors of $\mu = 8.5 \pm 0.5$ for $HST$ F110W, $\mu=8.3 \pm 0.3 $ for GTC $g$-band, $\mu=8.2^{+0.6}_{-0.8}$ 
for VLA, $\mu=9.2 \pm 0.8 $ for PdBI CO$\,(J = 5 \rightarrow 4$), and $\mu=9.2\pm0.5$ for the SMA dust continuum. 
The differential magnification appears to be small, since the bulk of the source emission of the LBG and SMG stands relatively 
far from the caustics without crossing them (see Figure~\ref{fig:offset}), and thus changes of magnification as a function of 
source plane position vary very little. We thus assume, for simplicity, lensing magnifications of $\mu_{\rm HST} = 8.5 \pm 0.5$ 
and $\mu_{\rm SMA} = 9.2 \pm 0.5$ to be the same in the spectral range in which the LBG and the SMG are well detected, respectively.

\begin{figure}[ht]
\includegraphics[width=0.46\textwidth]{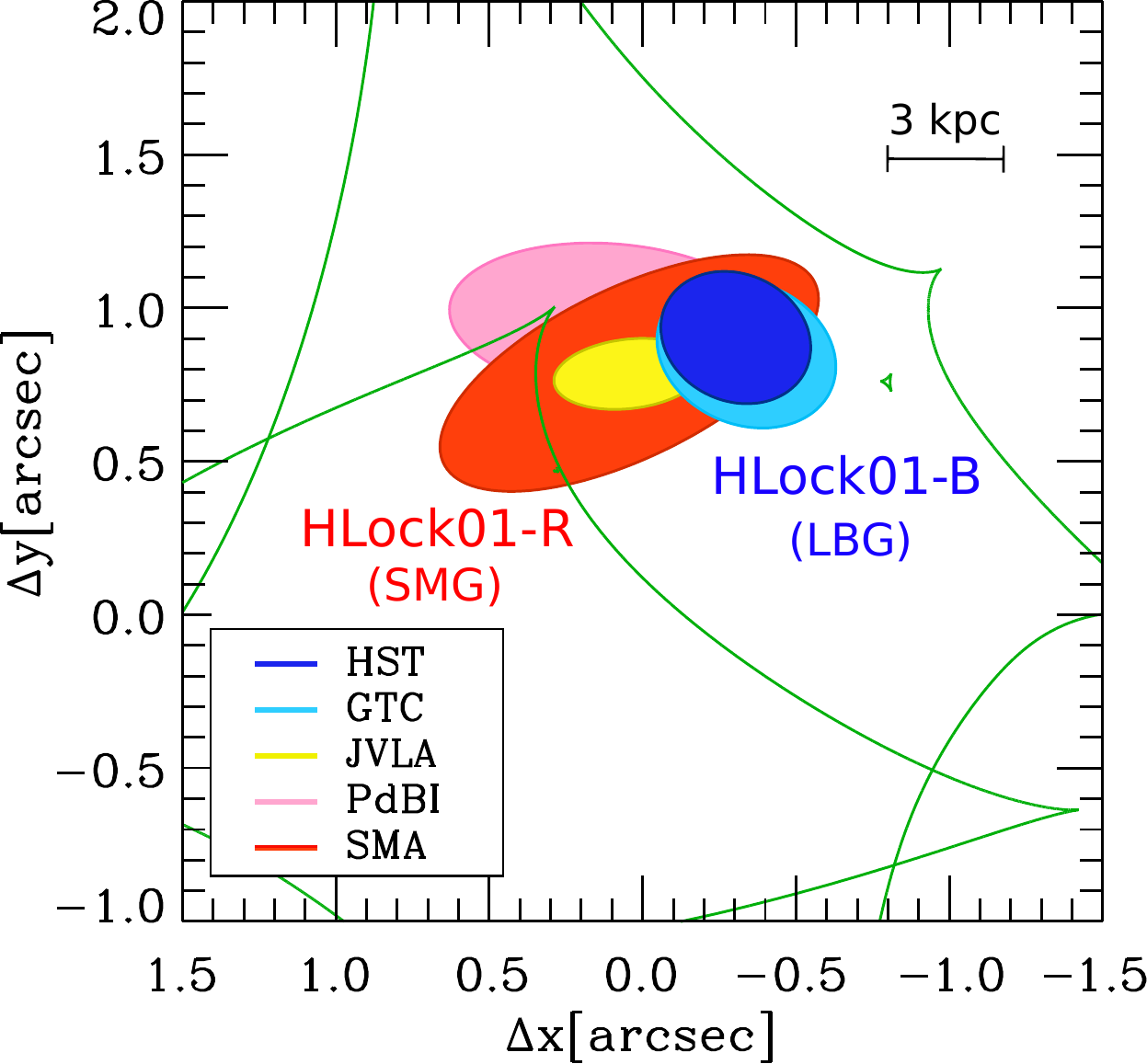}
\caption{Source plane reconstruction of HLock01. The posterior mean effective ellipses of the reconstructed components of HLock01-B 
(LBG) from $HST$ F110W and GTC $g$-band are represented by blue colored ellipses. Pink, yellow and red colored ellipses represent 
the effective radius of the reconstructed components of CO, VLA and dust emission associated with HLock01-R (SMG), respectively. 
A relative offset of $0\farcs42 \pm 0\farcs07$ is seen between $HST$ F110W and VLA, which corresponds to $3.3 \pm 0.6$ kpc. 
\label{fig:offset}}
\end{figure}

We firstly used the SED-fitting code {\sc Fast} \citep[Fitting and Assessment of Synthetic Templates;][]{kriek2009} to 
derive the stellar population properties of HLock01-B. 
Optical $U$, $g$, $R$, and $I$, and near-IR F110W and $K_{\rm s}$ flux measurements were used in this fit (see Appendix \ref{bb}). 
We excluded fluxes from IRAC from this fit, given the uncertainties of the contribution of HLock01-B in these bands. However, 
the 2.2 $\mu$m $K_{\rm s}$ band corresponds to rest-frame emission at $5600$ \AA, above the Balmer/$4000$ \AA \space break, 
which is sensitive to the age of the stellar population. 
We assume stellar population synthesis models of \cite{bruzual2003}, the \cite{chabrier2003} IMF, and an exponentially 
declining star-formation history ($\propto \rm e^{\rm -t / \tau}$).
We adopt a grid for the age of the stellar population, ranging from 20 Myr to the maximum age of the Universe at $z \simeq 2.95$, 
and star-formation histories with $\tau$ between 0.3 and 10 Gyr, both in steps of 0.1 dex. 
The attenuation curve of \cite{calzetti2000} was adopted, and the allowed $A_{V}$ range was $0-3$ mag in steps of 0.05 mag. 
We also fixed the metallicity to $Z / Z_{\odot} = 0.4$, the value measured in Section \ref{met} for the young O and B stars.
The best-fit model ($\chi^{2} / N_{\rm dof} = 1.6 / 3$) gives an intrinsic (i.e., corrected for the lensing magnification 
$\mu_{\rm HST} = 8.5 \pm 0.5$ and assumed to be the same in the spectral range in which the LBG is well detected) 
stellar mass log($M_{*}/ \rm M_{\odot}$)$ = 10.1_{-0.1}^{+0.3}$, and an attenuation of the stellar light of $A_{V} =  
0.84 _{-0.25}^{+0.12}$, with age log(age$_{M} /$yr$^{-1}) =  7.3_{-0.0}^{+0.6}$. Errors refer to $68\%$ confidence intervals 
derived using 500 Monte Carlo simulations. After correction for the lensing magnification, the star-formation rate of the best 
fit model is $\rm SFR = 710_{-420}^{+180}$ $\rm M_{\odot}$yr$^{-1}$.

\begin{figure*}[ht]
\centering
\includegraphics[width=160mm,scale=1]{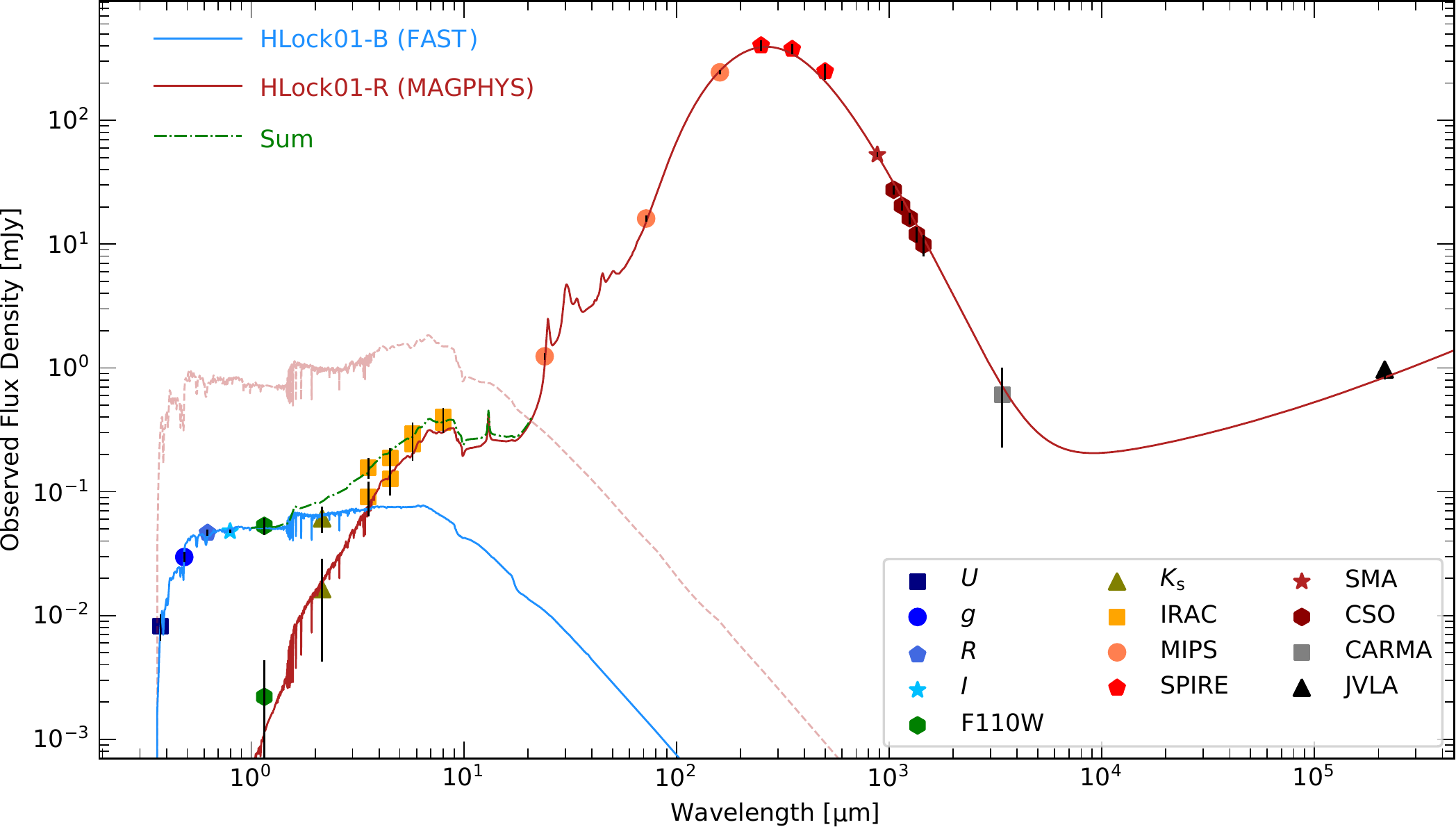}
\caption{Best-fit model of the spectral energy distribution of HLock01-B (LBG, blue solid line) using {\sc Fast} \citep{kriek2009}, 
and HLock01-R (SMG, red solid line) using {\sc Magphys} \citep{dacunha2008, dacunha2015}. 
The LBG fit uses photometry from CFHT $U$ to WHT $K_{\rm s}$ bands, whereas for the SMG the fit uses photometry from the 
rest-frame UV ($HST$ WFC3 F110W) to radio. We also used the estimated fluxes in the $Spitzer$ bands (3.6, 4.5, 5.8, and 8.0 $\mu$m) 
as explained in the text. The unattenuated SED of the SMG is plotted as a red dashed line. \label{fig:mag}}
\end{figure*}

We further performed a multi-band SED fit of HLock01-R using the high-$z$ extension of {\sc Magphys} \citep[Multi-wavelength 
Analysis of Galaxy Physical Properties;][]{dacunha2008, dacunha2015} to explore its SFR, and stellar and dust mass ($M_{\rm d}$). 
{\sc Magphys} uses the \cite{bruzual2003} stellar populations with a \cite{chabrier2003} IMF and assumes the attenuation model 
of \cite{dust2000}. We used the flux measurements from 1.1$\mu$m to radio. 
In the $Spitzer$/IRAC bands, we used the difference between the total fluxes (measured in Appendix \ref{bb}) and the expected 
flux of HLock01-B from the best-fit SED, as indicated in Table \ref{flux}.
The best-fit model (reduced $\chi^{2} = 3.1$) gives a lensing corrected ($\mu_{\rm SMA} = 9.2 \pm 0.5$) stellar mass 
log($M_{*}/ \rm M_{\odot}$)$ = 11.7_{-0.1}^{+0.2}$, a mass weighted age log$(age_{\rm M} /$yr$) =  8.6 \pm 0.1$, and a large 
attenuation $A_{V} =  4.26_{-0.10}^{+0.35}$ mag. We find a dust mass log($M_{\rm d}/ \rm M_{\odot}$)$ = 8.8 \pm 0.1$, with a dust 
temperature $T_{\rm d} = (53.6 \pm 0.2)$ K. The uncertainties are derived from the 16th and 84th percentiles.
The best-fit SED also yields an intrinsic total infrared luminosity $L_{\rm IR} = \rm (1.5 \pm 0.1) \times 10^{13} L_{\odot}$, 
which is defined as the luminosity from $8 - 1000$ $\mu$m in the rest frame. Using a Kennicutt relation \citep{kennicutt1998} 
with a Chabrier IMF \citep{chabrier2003}, the total infrared luminosity implies a star-formation rate of $\simeq 1500$ 
$\rm M_{\odot}$yr$^{-1}$. 
All values were corrected for the lensing magnification derived from the SMA $880\, \mu \rm m$ data ($\mu_{\rm SMA} = 9.2 \pm 0.5$), 
which for simplicity, we assume to be the same from the observed near-IR to submm bands for HLock01-R. 
 
We followed \cite{delvecchio2017} and \cite{miettinen2017} to look for a possible AGN contribution in HLock01-R. We use the 
three-component fitting code {\sc Sed3fit} \citep{berta2013}, which accounts simultaneously for stellar, dust, and AGN emission. 
However, the stellar and dust components of {\sc Sed3fit} use the model libraries of \cite{dacunha2008}, rather than the ones 
used in the new high-$z$ extension of {\sc Magphys} \citep{dacunha2015}, which are expected to be better suited for high-$z$ SMGs.
We found a poor fit to our data with a $\chi^{2} = 15.4$ for the best fit model, which is much higher than the $\chi^{2}$ 
of the standard {\sc Magphys} fit.  
Nevertheless, the mean AGN contribution to the total $L_{\rm IR}$ was found to be $ 1 \%$, and less than $ 20 \%$ in the IRAC bands. 
Our analysis does not completely exclude the possibility that HLock01-R may harbor an AGN \citep[see also][for a discussion of 
the presence of an AGN in HLock01]{conley2011, riechers2011, scott2011, magdis2014}, but its contribution to the total 
$L_{\rm IR}$ and $M_{*}$ is not substantial, as also pointed out by \cite{rigopoulou2018}.

Figure \ref{fig:mag} shows our best-fit SED models for HLock01-B ({\sc Fast}) and HLock01-R ({\sc Magphys}). Our results show that 
both galaxies are undergoing simultaneous episodes of star-formation activity (unobscured, nearly dust-free in HLock01-B, and 
dust-enshrouded star formation in HLock01-R), but they are physically very distinct. While HLock01-R is a very massive galaxy 
with an evolved stellar population, HLock01-B appears to be a young, lower-mass satellite of HLock01-R.

\section{Discussion}\label{discussion}

\subsection{Close merger or a large rotational disk?}\label{merger}

Our high S/N GTC/OSIRIS spectrum of the optically bright lensed images of HLock01 (HLock01-B) shows several well-defined UV 
photospheric absorption lines, for which we secured the systemic redshift $z_{\rm sys} = 2.9546 \pm 0.0004$. This value differs 
by $ -210$ km s$^{-1}$ from the redshift of HLock01-R measured from the molecular gas lines $z_{\rm CO} = 2.9574 \pm 0.0001$ 
\citep{riechers2011, scott2011}. 
A spatial offset of $3.3$ kpc (in projection) has also been found in the source plane between the bulk of the stars that emit 
at rest-frame UV/optical wavelengths (HLock01-B), and the molecular gas and dust distribution associated with the luminous far-IR 
emitting source HLock01-R. 

Although similar or even larger rotational velocities have been found in massive disk SMGs at high-$z$ 
\citep[e.g.,][]{carilli2010, daddi2010, jimenez2017, jones2017}, a scenario with HLock01-B being a dust-free region, that is part 
of a large rotational disk of HLock01-R, is unlikely. Such asymmetry in the dust distribution, with the lack of dust 
attenuation in HLock01-B, would be difficult to explain. Additionally, despite the large errors, the differences in 
the metallicity measured in the stars of HLock01-B and in the gas of the $Herschel$ SMG HLock01-R \citep{rigopoulou2018} 
suggest they are different galaxies with different enrichment histories.

Therefore, bringing together our GTC spectroscopic results and the complex velocity structure seen in the molecular 
gas reservoir in HLock01-R \citep{riechers2011}, we argue that HLock01 comprises two close but different sources forming a pair of 
merging galaxies (HLock01-B and HLock01-R) separated by $ 3.3$ kpc in projection. 
The merger scenario is also sustained by the broadness of the blueshifted ISM absorption lines seen in the 
spectrum of HLock01-B, suggesting highly turbulent gas likely produced by the close merger. 
While HLock01-R appears to be an evolved massive galaxy with a very large obscured star-formation rate, HLock01-B is a young 
lower-mass satellite galaxy with photometric properties similar to those of LBGs, yet undergoing a young burst ($\gtrsim 6$ Myr) 
of star formation, likely triggered by the gravitational interaction with the nearby massive SMG.

It is worth mentioning that without the gravitational lensing effect our results could easily be mistaken. 
Firstly, without the magnification in the apparent flux of HLock01-B, the systemic redshift, measured from faint stellar 
photospheric lines, would be difficult to obtain. In the absence of strong UV nebular emission lines, as is the case of 
HLock01-B, a significant continuum S/N is required to detect faint photospheric absorption lines. 
Secondly, the projected $ 3.3$ kpc spatial offset seen between the two different objects in the source plane would correspond 
to only $0.4^{\prime \prime}$ without the lensing distortion, which is challenging to observe due to the limitation on the 
spatial resolution and sensitivity of current instruments. 

It is thus no surprise that most high-$z$ close mergers ($\lesssim 10$ kpc projected separation) have been discovered 
through the gravitational lensing effect \citep[e.g.,][]{ivisonb2010, mackenzie2014, messias2014, rawle2014, wuyts2014, 
spilker2015, marrone2018}, with a few exceptions of unlensed, well separated, SMG-SMG, SMG-QSO or SMG-LBG bright 
interacting pairs resolved with interferometric observations \citep[e.g.,][]{ivison2002, ivison2008, smail2003, salome2012, 
oteo2016, lu2017, riechers2017}.

\subsection{Outflow/Inflowing gas}

Turning to the rest-frame UV spectral features, the internal kinematics of HLock01-B are very complex and different from 
what is observed in typical LBGs, showing two distinct components of the ISM. The blueshifted component of the ISM is 
centered at $v_{\rm ISM} \simeq -220$ kms$^{-1}$ relative to the stars, which we associate with galaxy-scale outflows of 
material via  stellar and supernova-driven winds, as seen in many other high-$z$ star-forming galaxies \citep[e.g.][]{shapley2003, 
steidel2010}. 
This component is stronger (i.e., larger equivalent widths) in high ionization lines, like Si {\sc iv} and C {\sc iv}, similar 
to what is found in other young, low-metallicity galaxies \citep[e.g.,][]{erb2010, james2014}. 
It is also detected in some low-ionization lines, but with a much broader profile ($\rm FWHM \simeq 900$ kms$^{-1}$) than in 
typical LBGs \citep[$\simeq 560$ kms$^{-1}$;][]{shapley2003, steidel2010}, extending over a large velocity range from 
approximately $-1000$ to $+600$ km s$^{-1}$.
We interpret that the broadness of the blueshifted absorption lines in HLock01-B is the result of a combination of strong winds of 
massive stars in the LBG and a complex velocity structure due to the close gravitational interaction with the massive SMG.

On the other hand, the redshifted component seen in all strong absorption lines (see Figures \ref{fig:low} and \ref{fig:gauss}) 
is highly unusual and not seen in the many high-$z$ galaxies studied \citep[e.g.,][]{shapley2003, steidel2010}. This component can 
be understood as gas apparently moving towards the young stars of HLock01-B, because the absorbing gas must lie in front of the LBG. 
We relate this component with the large column density of foreground neutral gas ($ N$(H~{\sc i})$=( 5.83 \pm 1.24) \times 10^{20}$ 
cm$^{-2}$) giving rise to the damped Ly$\alpha$ absorption seen in the spectrum of HLock01-B (see Section \ref{damp}).
The detection of this absorption in both low- and high-ionization ISM lines also suggests that the gas has a broad range of 
temperatures, from cold, mostly neutral (e.g., O {\sc i}) to warmer and ionized gas (e.g., Si {\sc iv}, and C {\sc iv}).

The origin and nature of the redshifted component seen in all strong absorption lines in the spectrum of HLock01-B is 
unclear, and with the available data we cannot arrive at a definite conclusion. 
Interpreting the redshifted component also depends strongly on the spatial location of the SMG and LBG along our line of sight. 
The relatively low dust attenuation in HLock01-B ($A_{\rm V} = 0.8_{-0.3}^{+0.1}$) may suggest that the SMG is located in the 
background, otherwise the stellar continuum of the LBG would be highly attenuated by the foreground dust content of HLock01-R 
(see Figure \ref{fig:offset}). Therefore, it is unlikely that the redshifted component is associated with outflows from HLock01-R 
or rotating gas in its disk seen from the background HLock01-B.

In this sense, the redshifted component could be associated with a dwarf galaxy or a damped Ly$\alpha$ system falling towards HLock01. 
Gas ejected by a previous episode of star formation or AGN activity in HLock01 would not easily escape its intense gravitational pull. 
If cooled enough, these reservoirs of gas would provide additional fuel to prolong the star-formation activity \citep[e.g.,][]{dave2011, 
hopkins2014, narayanan2015, wang2015, emonts2016}.
Assuming that the gas is dynamically linked and collapsing towards HLock01, it is more likely that the gas is falling into the massive 
SMG with $v \simeq +170$ km s$^{-1}$, instead of falling to HLock01-B with $v \simeq +370$ km s$^{-1}$, which seems too high for a 
$10^{10}$ $\rm M_{\odot}$ galaxy.

Evidence of accretion of cool, metal-enriched gas has been found in only a few spectra of star-forming galaxies at moderately low-$z$ 
\citep[e.g.,][]{sato2009, coil2011, rubin2012, martin2012}, and is more elusive at high-$z$ \citep[e.g.,][]{bouche2013, wiseman2017} 
due to the faintness of individual high-$z$ galaxies. 
Some authors suggest that the low detection rate of infalling gas is due to the geometry and alignment of the streams, which can 
only be detected in absorption if favorably aligned with our line of sight \citep{kimm2011, martin2012}. 
Nevertheless, accretion of cold gas, either in the form of cold flows, mergers or recycled gas from stellar feedback, 
plays an important role in star-formation histories and galaxy growth. \\

\subsection{Extended gas reservoir?}\label{reservoir}

We reported an unusual absorption line at $\simeq 5281$ \AA \space in the red-wing of a bright Ly$\alpha$ emission at $z \simeq 3.327$, 
associated with an object $14^{\prime \prime}$ SW of HLock01 (see Appendix \ref{lensing}). 
This absorption is consistent with C {\sc ii} $\lambda \lambda$1334 at $z = 2.9574 \pm 0.0008$, very close by $\Delta v = (-8 \pm 76)$ 
km s$^{-1}$ to the redshift of HLock01-R measured from the molecular gas ($z_{\rm CO} = 2.9574 \pm 0.0001$). 
The limited spectral coverage of the Ly$\alpha$ emission line and the faintness of the continuum associated with the $z \simeq 3.327$ 
galaxy does not let us unambiguously confirm if the absorption line is C {\sc ii} at the redshift of HLock01-R, or a different 
absorption line system at a lower redshift. If related with HLock01-R, it may suggest a substantial gas reservoir in the halo at 
an impact parameter of $110$ kpc. 
It is worth noting that \cite{fu2016} used QSO absorption line spectroscopy in three high-$z$ SMG-QSO close pairs, with the QSO at a 
larger redshift than the SMG, to probe the circumgalactic medium (CGM) at similar impact parameters. However, they did not find 
evidence of optically thick H {\sc i} gas or strong neutral absorbers in the CGM. Our results suggest that at least massive SMGs, 
such as HLock01-R, may have prominent cool gas reservoirs in their halos, that could fuel a prolonged star formation phase.

\subsection{Physical Properties}

The intrinsic physical properties (corrected for lensing magnification) derived from multiband SED fitting reveal that HLock01-R is 
a far-IR luminous SMG \citep[as already discussed in earlier papers, e.g.,][]{conley2011, wardlow2013}, with an ongoing $\rm SFR 
\simeq 1500$ $\rm M_{\odot}$yr$^{-1}$, but yet heavily obscured at short wavelengths. 
Our new analysis reveals a large, highly obscured stellar mass, similar to the most massive and extreme SMGs during the peak of 
star formation \citep[e.g.,][]{hainline2011, ma2015, schinnerer2016, miettinen2017, nayyeri2017}.
However, as already discussed in \cite{conley2011} and \cite{wardlow2013}, HLock01-R has a moderately low $q_{\rm IR} = 1.8 \pm 0.4$ 
(which is the logarithmic ratio of $L_{\rm IR}$ and the rest-frame 1.4GHz flux density), compared with the mean value for HerMES 
sources \citep[$q_{\rm IR} = 2.40 \pm 0.12$;][]{ivison2010}.
This may indicate a hidden, radio emitting AGN, but from our SED analysis in Section \ref{sed}, we have shown that if HLock01-R 
harbors an AGN, its contribution to the total $L_{\rm IR}$ and SFR is modest ($\simeq 1 \%$), as also noted by \cite{rigopoulou2018}, 
and even assuming a maximum of $20 \%$ of an AGN contribution to the IRAC fluxes, the stellar mass of HLock01-R will be lower 
only by 0.1 dex, which is within our measurement errors. 
Moreover, \cite{hayward2015} have also shown that the physical properties derived using {\sc Magphys} are robust even when 
the AGN contributes $25\%$ of the total UV to IR luminosity. 
Our deep GTC/OSIRIS rest-frame UV spectroscopy does not show any line or continuum emission at the positions of the lensed images of 
HLock01-R, as some of them are included in the regions covered by our long-slit spectra (see Figure \ref{fig:rgb}, left panel). 
However, follow-up observations are required to constrain the presence of an AGN in HLock01-R.
Nevertheless, even assuming a small AGN contribution, our results show that HLock01-R has already formed the majority of its stellar 
content, with a gas mass fraction of $f_{\rm gas} \equiv M_{\rm gas} / M_{*} = 0.07 \pm 0.02$ \citep[for $M_{\rm gas}=3.3 \times 
10^{10}$~$\rm M_{\odot}$, as measured in][]{riechers2011}, a specific star formation rate $\rm sSFR \equiv \rm SFR / M_{*} = 
2.76_{-0.85}^{1.22}$~Gyr$^{-1}$, and a depletion time scale ($\tau_{\rm d} \equiv M_{\rm gas} / \rm SFR$) of only $22$ Myr (assuming 
no gas input). Moreover, it is plausible that additional gas input from the ongoing merger and inflows of material from a substantial 
gas reservoir in the halo will extend the starburst phase of HLock01 for a prolonged time, becoming an even more massive elliptical 
galaxy in the local Universe.

On the other hand, HLock01-B appears to be a young, lower-stellar mass galaxy with very different properties than HLock01-R. 
In the optical, it is one of the brightest gravitationally lensed high-$z$ star-forming galaxies known so far 
\citep[e.g.,][]{yee1996, allam2007, belokurov2007, smail2007, lin2009, wuyts2010, bayliss2011, dahle2016, marques2017}, with an apparent 
total magnitude of $R = 19.73 \pm 0.01$. 
Even after accounting for the magnification produced by the lensing group of galaxies ($\mu_{\rm HST} = 8.5 \pm 0.5$), it is still very 
luminous in the rest-frame UV with an absolute magnitude $M_{\rm UV} = -23.4$, two and a half magnitudes more luminous than typical 
LBGs ($L^{*}_{\rm UV}$) at a similar redshift \citep{reddy2009}. 
The stellar mass and SFR derived in Section \ref{sed} yield a specific star formation rate of $55$ Gyr$^{-1}$ well above the main 
sequence at that redshift \citep[e.g.,][]{mannucci2009, magdis2010b, alvarez2016}. 
The nature and properties of this kind of UV ultra-luminous galaxies (i.e., $M_{\rm UV} \lesssim -23$) are still poorly understood, 
given the lack of examples reported in the literature \citep[e.g.,][]{allam2007, bian2012, lefevre2013, ono2017, marques2017}. 
This is due in part to the fact that this kind of galaxy is extremely rare, and finding them requires wider-field surveys with deep, 
multi-band observations. 
The high UV luminosities also place these sources in the transition between luminous galaxies and faint AGNs, needing either extensive 
multi-wavelength imaging (e.g., X-ray, mid-IR, and radio) or spectroscopic follow-up. 
However, we stress that the unusual kinematics of the ISM and its high UV luminosity and SFR are not representative of the 
$z \sim 3$ LBG population. Our results suggest that the gravitational interaction with the massive SMG may have triggered the 
larger UV luminosity and SFR. The interaction may also be the origin of the large obscured SFR and far-IR luminosity in the SMG. 
Despite this, HLock01-B shares many of its properties with the population of $z \sim 3$ LBGs. Its UV colors, $(G - R) \simeq 0.5$, 
and $(U - G) \simeq 1.4$, are consistent with the standard color selection criteria of $z \sim 3$ LBGs \citep{steidel1996, steidel2003}. 
However, HLock01-B presents $(R - K) = 0.29 \pm 0.07$, bluer than typical $z \sim 3$ LBGs \citep[$(R - K) \simeq 1.0$;][]{shapley2001} 
and Ly$\alpha$ emitting galaxies \citep[$(R - K) \simeq 0.4$;][]{ono2010}. 

Table \ref{tab6} summarizes the main physical properties of both components of HLock01.

\begin{table}[ht]
\begin{center}
\tabletypesize{\scriptsize}
\caption{De-magnified Physical Properties of HLock01. \label{tab6}}
\begin{tabular}{c c c c }
\hline \hline
\smallskip
\smallskip
Quantity & HLock01-B & HLock01-R & Unit  \\
\hline
$z$  & $2.9546 \pm 0.0004$ & $2.9574 \pm 0.0001$ & -- \\
$M_{*}$ &  $10.1_{-0.1}^{+0.3}$ & $11.7_{-0.1}^{+0.2}$ & log($\rm M_{\odot}$) \\
Age    &  $7.3_{-0.0}^{+0.6}$ &  $8.6 \pm 0.1$ &  log($\rm yr$) \\
$A_{\rm V}$ & $0.8_{-0.3}^{+0.1}$ & $4.3_{-0.1}^{+0.4}$ & mag \\
SFR  & $710_{-420}^{+180}$  & $1500 \pm 200$ & $\rm M_{\odot}$yr$^{-1}$ \\
sSFR & $55_{-27}^{+14}$ & $2.76_{-0.85}^{+1.22} $ & Gyr$^{-1}$ \\
$Z$ & $0.4 \pm 0.1$ & $0.8 \pm 0.2$ & $Z_{\odot}$ \\ 
\hline 
\end{tabular}
\\
\end{center}
\end{table}

\section{Summary and Conclusions}\label{conclusion}

We have presented a detailed study of HLock01, one of the first gravitational lensed sources discovered in the HerMES survey. 
Unlike other SMGs, HLock01 is apparently very bright in all observed spectral bands, even in the optical.
It is magnified by a factor of around $9$ by a galaxy group-scale dark matter halo at $z = 0.645$ and comprises four images in the 
observed plane. 
We have used OSIRIS on the GTC to secure a high S/N ($\simeq 30$) rest-frame UV spectrum of the optically bright lensed 
images of HLock01, with an intermediate-resolution ($\simeq 180$ km s$^{-1}$), covering the wavelength interval $1150 - 1950$ 
\AA \space in the rest-frame. 
From the analysis of these data together with other existing observations of HLock01, we arrive at the following main results.
\\
\\
1. We measured the systemic redshift of the optically bright lensed images of HLock01 (HLock01-B) 
$z_{\rm sys} = 2.9546 \pm 0.0004$ using weak stellar photospheric lines. 
This value is offset by $-210$ km s$^{-1}$ from the redshift measured previously from the molecular gas lines 
$z_{\rm CO} = 2.9574 \pm 0.0001$ associated with the luminous far-IR source of HLock01 (HLock01-R). 
Our results show that the dust-obscured, far-IR emitting source HLock01-R, and the optically bright source HLock01-B, are 
most likely different galaxies undergoing a close merger or an interacting pair separated by only $3.3$ kpc in projection.  
\\
\\
2. We find a stellar metallicity for the stars in HLock01-B $Z_{\rm stars} \simeq 0.4 Z_{\odot}$ based on two independent methods: 
blends of stellar photospheric lines; and P-Cygni profiles from the most luminous O and B stars. This value differs slightly from 
that measured for the gas in HLock01-R ($0.6 < Z_{\rm gas} / Z_{\odot} < 1.0$), based on far-IR fine-structure line ratios.
A young ($\gtrsim 6$ Myr) starburst model with a Salpeter IMF, stellar masses from 1 to 100 $\rm M_{\odot}$, and an LMC/SMC metallicity 
explains well the properties of the high-ionization lines in HLock01-B.  
\\
\\
3. The interstellar absorption lines in the spectrum of HLock01-B exhibit two distinct components. 
One is blueshifted by $-220$ km s$^{-1}$ relative to the 
stars of HLock01-B, which we associate with galaxy-scale outflows via stellar and supernovae driven winds. 
However, it also shows a broader profile ($\rm FWHM \simeq 900$ km s$^{-1}$) than in most star-forming galaxies at 
$z = 2- 3$, indicating highly turbulent kinematics of the outflowing gas likely due to the close merger. 
This component is stronger in high ionization lines, suggesting that the gas is mostly ionized or the neutral 
gas has a lower covering factor than the ionized gas. 

Another absorption component is seen in the spectrum of HLock01-B, but is redshifted relative to either HLock01-B and HLock01-R 
by $+370$ and $+170$ km s$^{-1}$, respectively, which can be understood as gas moving towards both galaxies. 
We relate this component with the strong damped Ly$\alpha$ line seen in HLock01-B, with a column density of 
$ N$(H~{\sc i})$=( 5.83 \pm 1.24) \times 10^{20}$ cm$^{-2}$. 
Although with the available data we cannot arrive at a definitive conclusion on its nature and origin, we interpret this 
absorption feature as gas falling towards HLock01-R, which is more massive, with $v \simeq 170$ km s$^{-1}$, but viewed in 
absorption along a favorable line of sight towards HLock01-B. 
This component is detected in both low- and high-ionization interstellar lines, but with slightly different absorption 
line profiles, suggesting that the gas has a broad range of temperatures, and possibly different origin. 
\\
\\
4. We detected an unusual absorption line in the red wing of the bright Ly$\alpha$ emission at $z \simeq 3.327$ at 
$14^{\prime \prime}$ SW from the lensing galaxy G1, that we tentatively associate with C {\sc ii} $\lambda \lambda$1334 at 
the redshift of HLock01-R. If this absorption is related with HLock01-R and not with an absorbing system at a different 
redshift, it indicates a substantial gas reservoir in the halo of HLock01 at a projected distance of $110$ kpc. 
Additionally, we report a broad absorption line QSO at a projected distance of $ 2.5$ Mpc from HLock01, with a 
redshift very close to HLock01-R ($\Delta v \simeq 300$ km s$^{-1}$).  
\\
\\
5. Our revised SED fitting with two different galaxies, one very bright in the optical and the other 
in the far-IR, implies that both are physically very distinct. 
HLock01-B appears to be a young, lower-mass satellite galaxy of HLock01-R, undergoing an intense episode 
of star formation activity likely triggered by the interaction. 
HLock01-R shows an already evolved stellar population, and its high stellar mass in combination with the 
low gas fraction suggests that the SMG has already assembled most of its stellar mass. However, additional 
gas input from the satellite galaxy HLock01-B and from the reservoir of gas around HLock01-R may extend the starburst 
phase of the SMG, eventually forming one of the most massive galaxies in the local Universe.
\\
\\

\acknowledgments

We would like to thank the anonymous referee for their suggestion which significantly improved the clarity of this paper. 
We also thank Alice Shapley for allowing us to use their rest-frame UV composite spectrum of LBGs and Helmut Dannerbauer 
for useful discussions. Based on observations made with the Gran Telescopio Canarias (GTC) and with the William Herschel Telescope 
(WHT), both installed in the Spanish Observatorio del Roque de los Muchachos of the Instituto de Astrof\'isica de Canarias, on the 
island of La Palma. We thank the GTC and WHT staff for their help with the observations. 
R.M.C. acknowledges Fundaci\'on La Caixa for the financial support received in the form of a PhD contract. 
R.M.C., I.P.F., P.M.N., and C.J.A. acknowledge support from the Spanish Ministerio de Economia y Competitividad (MINECO) under 
grant number ESP2015-65597-C4-4-R. 
Y.S. has been partially supported by the 973 program (No.~2015CB857003) and the National Natural Science Foundation of 
China (NSFC) under grant numbers 11603032 and 11333008.
J.L.W. gratefully acknowledges an STFC Ernest Rutherford Fellowship and additional support from STFC (ST/P000541/1).

SPIRE has been developed by a consortium of institutes led by Cardiff University (UK) and including: Univ.~Lethbridge (Canada); NAOC
(China); CEA, LAM (France); IFSI, Univ.~Padua (Italy); IAC (Spain); Stockholm Observatory (Sweden); Imperial College London, RAL, 
UCL-MSSL, UKATC, Univ. Sussex (UK); and Caltech, JPL, NHSC, Univ. Colorado (USA). This development has been supported by national funding 
agencies: CSA (Canada); NAOC (China); CEA, CNES, CNRS (France); ASI (Italy); MCINN (Spain); SNSB (Sweden); STFC, UKSA (UK); and NASA (USA). 

Funding for SDSS-III has been provided by the Alfred P. Sloan Foundation, the Participating Institutions, the National Science 
Foundation, and the U.S. Department of Energy Office of Science. The SDSS-III web site is \url{http://www.sdss3.org/}.
SDSS-III is managed by the Astrophysical Research Consortium for the Participating Institutions of the SDSS-III Collaboration, 
including the University of Arizona, the Brazilian Participation Group, Brookhaven National Laboratory, Carnegie Mellon 
University, University of Florida, the French Participation Group, the German Participation Group, Harvard University, the 
Instituto de Astrofisica de Canarias, the Michigan State/Notre Dame/JINA Participation Group, Johns Hopkins University, Lawrence 
Berkeley National Laboratory, Max Planck Institute for Astrophysics, Max Planck Institute for Extraterrestrial Physics, New Mexico 
State University, New York University, Ohio State University, Pennsylvania State University, University of Portsmouth, Princeton 
University, the Spanish Participation Group, University of Tokyo, University of Utah, Vanderbilt University, University of 
Virginia, University of Washington, and Yale University.

\vspace{5mm}
\facilities{GTC (OSIRIS), $HST$ (WFC3), WHT (LIRIS), $Spitzer$ (IRAC), $Herschel$ (SPIRE), SMA, VLA. }

\begin{appendix}
\section{Enviroment of HLock01}\label{lensing}

\begin{figure*}[ht]
\centering
\includegraphics[width=180mm,scale=1]{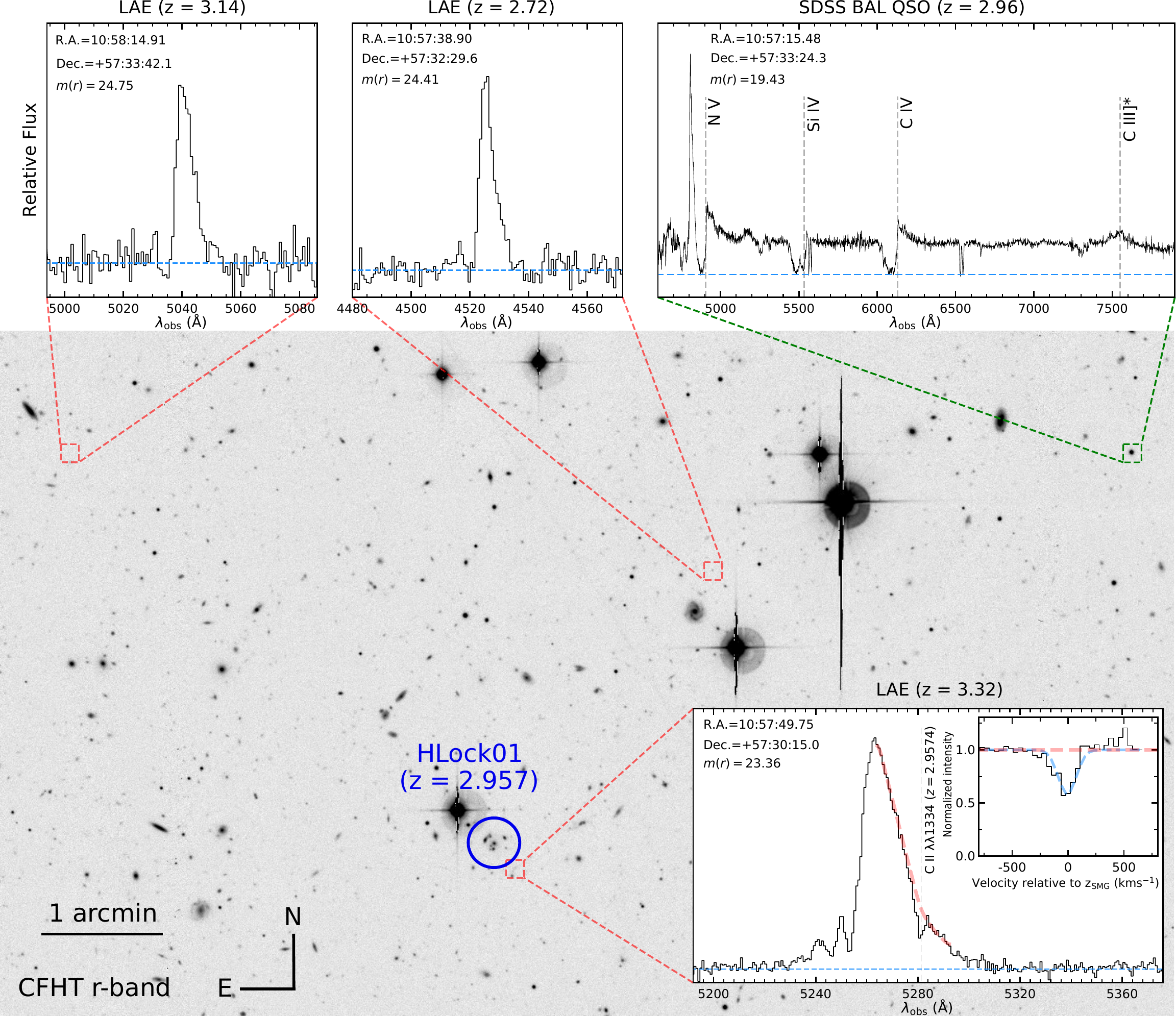}
\caption{CFHT $R$-band showing the spatial distribution of the $z \simeq 2.96$ BAL QSO SDSS J105715.48+573324.3 
(top right corner, green dashed square), and three high-redshift galaxies (LAEs) serendipitously detected in two of our GTC long-slit 
spectra (red dashed squares). The position of HLock01 is marked in blue. The redshift of the Ly$\alpha$ lines, as well as the 
coordinates and $R$-band magnitudes of the associated objects, are labeled in each panel. In particular, the 
$z = 3.327$ Ly$\alpha$ emission (bottom right corner) shows an unresolved absorption line at $5281$ \AA \space in its red wing. 
This absorption is consistent with the low-ionization line C {\sc ii} $\lambda \lambda$1334 at $z = 2.9574 \pm 0.0008$ (inset panel), 
very close ($\Delta v = - 8 \pm 76$ km s$^{-1}$) to the redshift of HLock01-R measured from the molecular gas \citep[$z_{\rm CO} = 
2.9574 \pm 0.0001$;][]{riechers2011, scott2011}. 
\label{fig:high_laes}}
\end{figure*}

We serendipitously detected in two of our $1.2^{\prime \prime}$-wide GTC long-slit spectra ($\rm PA= -39^{\circ}.5$ and $44^{\circ}$) 
three strong, asymmetric lines that we interpret as Ly$\alpha$ emission at  $z = 2.721 \pm 0.001$, $z = 3.145 \pm 0.001$, 
and $z = 3.327 \pm 0.001$, at $2.37^{\prime}$, $4.35^{\prime}$, and $14^{\prime \prime}$ from the lensing galaxy G1, respectively.  
Figure \ref{fig:high_laes} shows the profiles of the Ly$\alpha$ emission, as well as the coordinates and magnitudes of the associated 
objects seen in CFHT $R$-band data. 

In particular, the Ly$\alpha$ emission at $z=3.327$ shows a broad profile ($\rm FWHM \simeq 1000$~km~s$^{-1}$, after accounting 
for the instrumental broadening), and has an observed flux of $F_{\rm Ly\alpha} = (1.38 \pm 0.2) \times 10^{-15}$ erg s$^{-1}$ cm$^{-2}$. 
A more detailed analysis of this object will be presented in Marques-Chaves et al. (in prep.), based on recent observations with GTC. 
We noticed that	the red-wing of the Ly$\alpha$ emission shows an unusual and unresolved ($\rm FWHM < 180$~km~s$^{-1}$) 
absorption line at $5281.2 \pm 0.9$ \AA \space (see Figure \ref{fig:high_laes}, bottom right corner), which is not related to this galaxy. 
This could be an absorption system at any lower redshift, but surprisingly it is consistent with C {\sc ii} $\lambda \lambda$1334 
absorption at $z = 2.9574 \pm 0.0008$, very close ($\Delta v = -8 \pm 76 $ km s$^{-1}$) to the redshift of HLock01-R measured from the molecular gas 
\citep[$z_{\rm CO} = 2.9574 \pm 0.0001$;][]{riechers2011, scott2011}. If this absorption is physically related with HLock01, it may 
suggest a substantial gas reservoir in its halo, at an impact parameter $b = 110$ kpc.

We also report on a $z = 2.961$ quasar, SDSS J105715.48+573324.3, at $5.5^{\prime}$ NW from HLock01 (see Figure 
\ref{fig:high_laes} at the top right corner). This object was cataloged as a broad absorption line quasar (BAL QSO) by 
\cite{trump2006} from the third edition of the Sloan Digital Sky Survey \citep[SDSS:][]{york2000} Quasar Catalog 
\citep{schneider2005}. The redshift of this BAL QSO is very close to the one of HLock01-R ($\Delta v \simeq 300$~km~s$^{-1}$) and 
is located at a projected distance of $2.5$ Mpc.

\section{Broad-band photometry}\label{bb}

In addition to the photometry presented in \cite{conley2011} and \cite{wardlow2013}, we also performed photometry on the new 
imaging data. These measurements are summarized in Table \ref{flux}.

\begin{table*}[ht]
\begin{center}
\tabletypesize{\scriptsize}
\caption{Photometry of HLock01. \label{flux}}
\begin{tabular}{c c c c c c}
\hline \hline
\smallskip
\smallskip
Telescope/Detector & $\lambda$  & HLock01-B$^{\rm a}$ & HLock01-R$^{\rm a}$ & Units & References \\
 & ($\mu$m) & (LBG) & (SMG) &  & \\
\hline
CFHT/MEGACAM ($U$) & 0.38 & $8.2 \pm 0.6$ & --- & $\mu$Jy & \\
GTC/OSIRIS ($g$) & 0.48 & $29.7 \pm 0.8$ & ---  & $\mu$Jy & \\
INT/WFC ($R$) & 0.63 & $46.5 \pm 0.6$ & ---  & $\mu$Jy & \cite{conley2011} \\
Subaru/SuprimeCam ($I$) & 0.76 & $47.9 \pm 0.04$ & ---  & $\mu$Jy & \cite{conley2011} \\
$HST$/F110W & 1.16 & $53.5 \pm 2.5$ &  $2.3 \pm 0.6$ & $\mu$Jy & \\
WHT/LIRIS ($K_{\rm s}$) & 2.20 & $60.8 \pm 4.7^{\rm b}$ & $16.3 \pm 3.8^{\rm b}$ & $\mu$Jy & \\
$Spitzer$/IRAC (I1) & 3.6 & $74 \pm 9^{\rm c} $ & $82 \pm 9^{\rm d}$ & $\mu$Jy & \\
$Spitzer$/IRAC (I2) & 4.5 & $75 \pm 11^{\rm c}$ & $113 \pm 11^{\rm d}$ & $\mu$Jy & \\
$Spitzer$/IRAC (I3) & 5.8 & $76 \pm 20^{\rm c}$ & $219 \pm 20^{\rm d}$ & $\mu$Jy & \\
$Spitzer$/IRAC (I4) & 8.0 & $63 \pm 20^{\rm c}$ & $341 \pm 20^{\rm d}$ & $\mu$Jy & \\
$Spitzer$/MIPS & 24 & ---  & $1.24 \pm 0.02$ & mJy & \cite{wardlow2013} \\
$Spitzer$/MIPS & 72 & ---  & $16.1 \pm 0.3$ & mJy & \cite{wardlow2013} \\
$Spitzer$/MIPS & 160 & ---  & $244.4 \pm 1.4$ & mJy & \cite{wardlow2013} \\
$Herschel$/SPIRE & 250 & ---  & $403 \pm 7$ & mJy & \cite{wardlow2013} \\
$Herschel$/SPIRE & 350 & ---  & $377 \pm 10$ & mJy & \cite{wardlow2013} \\
$Herschel$/SPIRE & 510 & ---  & $249 \pm 7$ & mJy & \cite{wardlow2013} \\
SMA & 880 & ---  & $52.8 \pm 0.5$ & mJy & \cite{conley2011} \\
CSO/Z-Spec & $1000 - 1100$ & ---  & $27.5 \pm 0.6$ & mJy & \cite{conley2011} \\
CSO/Z-Spec & $1100 - 1200$ & ---  & $20.4 \pm 0.5$ & mJy & \cite{conley2011} \\
CSO/Z-Spec & $1200 - 1300$ & ---  & $16.2 \pm 0.5$ & mJy & \cite{conley2011} \\
CSO/Z-Spec & $1300 - 1400$ & ---  & $12.0 \pm 0.5$ & mJy & \cite{conley2011} \\
CSO/Z-Spec & $1400 - 1500$ & ---  & $9.9 \pm 0.6$ & mJy & \cite{conley2011} \\
CARMA & 3400 & ---  & $0.61 \pm 0.19$ & mJy & \cite{conley2011} \\
VLA & 214000 & ---  & $0.97 \pm 0.05$ & mJy & \cite{wardlow2013} \\
\hline  
\end{tabular}
\\
\end{center}
\textsc{      \bf{Notes.}} \\
$^{\rm a}$ Total flux densities of the four lensed components, uncorrected for lensing magnification. \\
$^{\rm b}$ Obtained by modeling the light profiles using {\sc galfit}. \\
$^{\rm c}$ Expected $Spitzer$/IRAC fluxes of HLock01-B, extrapolated from the best fit SED using flux measurements from 0.38 to 2.20 
$\mu$m (see Section \ref{sed}). \\
$^{\rm d}$ Refers to the difference between the total flux densities measured in Appendix \ref{bb} (156, 188, 295, and 404 $\mu$Jy 
for the $Spitzer$/IRAC bands I1, I2, I3, and I4, respectively) and the expected flux densities of HLock01-B from the best-fit SED. \\
 \\
\end{table*}

We use aperture photometry in the $U$ band from the corresponding CFHT/MEGACAM catalog, which contains detections of all four lensed images, 
since the light contamination from the red, lensing galaxies is negligible in $U$-band. 
For GTC $g$-band, we applied aperture photometry on the lensed images A, C, and D. 
We exclude photometry of the lensed image B, as it is strongly blended with the lens galaxy G4, but we use the lens 
model to correct for the omitted light from image B (a roughly $15 \%$ correction).

\begin{figure*}[ht]
\centering
\includegraphics[width=180mm,scale=1]{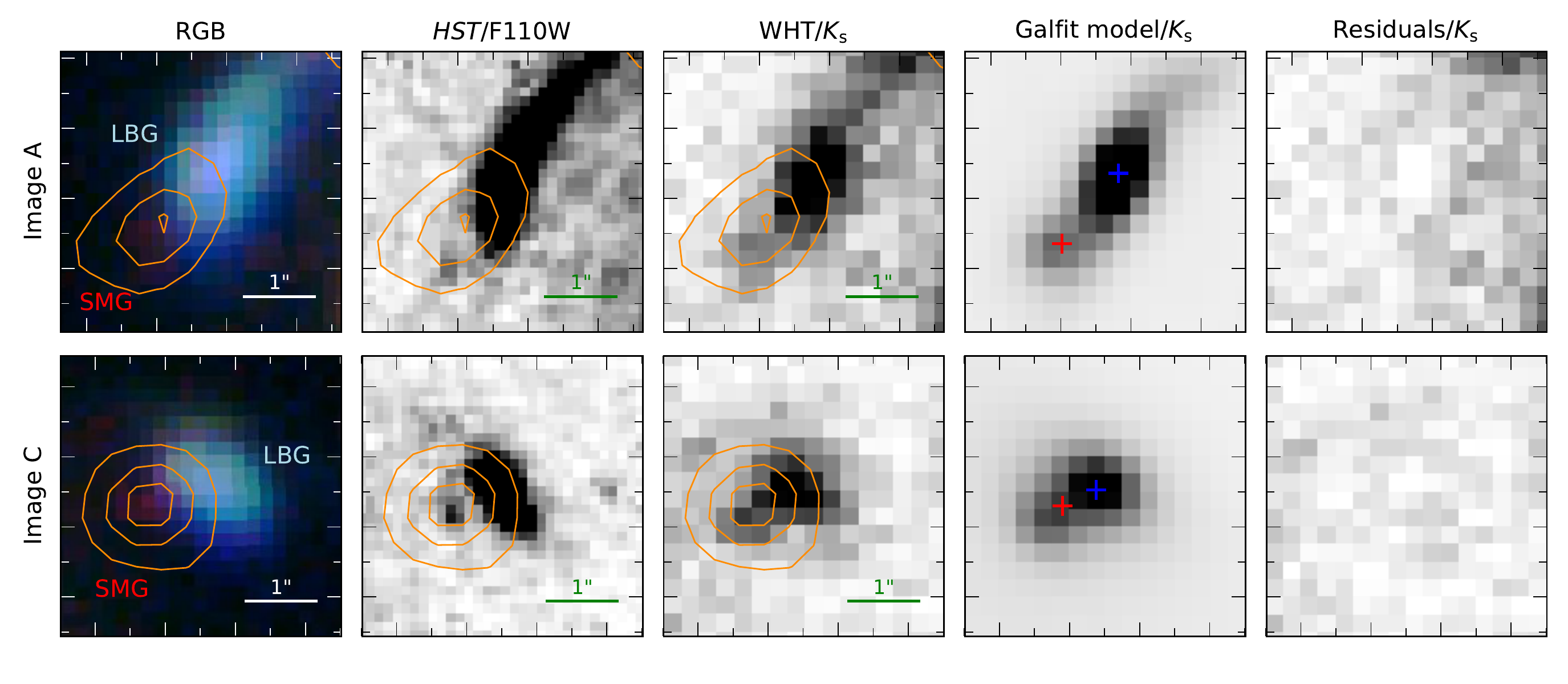}
\caption{$4'' \times 4''$ cutouts of the lensed images A (upper panels) and C (lower panels) of HLock01. 
From left to right we show: a $g$, $I$, and $K_{\rm s}$ color image; high-resolution $HST$/WFC3 F110W; near-IR WHT/$K_{\rm s}$; 
{\sc galfit} model of the $K_{\rm s}$ data; and finally the resulting residuals after subtracting the $K_{\rm s}$ {\sc galfit} model.  
1.4GHz VLA contours are in orange and correspond to the positions of two of the lensed images of the $Herschel$ SMG. 
Blue and red crosses mark, respectively, the positions of the LBG and the SMG in our {\sc galfit} model. \label{fig:galfit}}
\end{figure*}

Despite the short exposure time of the $HST$ F110W data, faint emission is seen close to the radio and submm lensed images A and C (see 
Figure \ref{fig:galfit}). 
$K$-band imaging from WHT/LIRIS and NIRC2/Keck-II \citep[the latter discussed in][]{gavazzi2011, calanog2014} also reveal 
faint emission at these positions. 
The red colors from the optical to 2.2$\mu$m imaging seen in Figure \ref{fig:galfit} (left panel) and Table \ref{flux} suggest that this 
faint emission corresponds to the obscured rest-frame UV and optical light of the $Herschel$ SMG. 
Associating this faint emission to HLock01-R, we measure the flux in a small aperture ($0.77^{\prime \prime}$ 
diameter) at the position of the faint near-IR source detected close to the radio and submm lensed image C. 
For HLock01-B, we use larger apertures ($2^{\prime \prime}-3^{\prime \prime}$) on the lensed images A, C, and D, 
and then subtract the contribution of HLock01-R, which in any case is less than $5 \%$.
The lens model was used again to add the light from the lensed image B. 

Near-IR WHT/$K_{\rm s}$ photometry of the individual components (SMG and LBG of HLock01) was obtained 
after modeling the light distribution of each component in the lensed images A and C, using the two dimensional fitting program 
{\sc galfit} \citep{peng2002, peng2010}. 
We use S\'ersic profiles centered at the centroids of the $HST$/F110W emission, allowing only one pixel freedom 
($\simeq 0.254^{\prime \prime}$). A nearby star was chosen as a point-spread function (PSF) model. 
Note that we only perform the fit in the lensed images A and C, the only ones that show detections of the faint, obscured 
counterparts of the SMG~\footnote{This can be explained through the lensed images A and C being less affected 
by foreground contamination and their lensing magnifications are higher than the ones of images B and D \citep[see][]{gavazzi2011}.}.
The lens model of \cite{gavazzi2011} was used again to add the light from the lensed images B and C. 
Figure \ref{fig:galfit} shows our {\sc galfit} model, as well as the resulting residuals after subtracting the $K_{\rm s}$ {\sc galfit} model.

This field has been observed by the SWIRE survey \citep[][]{lonsdale2003} in the cryogenic phase of $Spitzer$ and to deeper levels in the two 
first bands of IRAC (3.6 and 4.5$\mu$m) by the SERVS survey \citep[][]{mauduit2012} in the post-cryogenic phase. 
In the $Spitzer$ Enhanced Imaging Products (SEIP) catalog\footnote{\url{http://irsa.ipac.caltech.edu/data/SPITZER/Enhanced/SEIP/}} of this 
area (based on the SWIRE data) the individual lensed images A and C are the only ones resolved and detected. 
However their fluxes appear relatively large and inconsistent ($f_{\rm A} / f_{\rm C} \lesssim 1$) with the expected values from the 
individual magnifications provided by the lens model of \cite{gavazzi2011} ($f_{\rm A} / f_{\rm C} \simeq 1.7$). 
Given the limited spatial resolution of IRAC ($\simeq 2^{\prime \prime}$), the cataloged fluxes of the lensed images A and C are 
likely contaminated by foreground light, mainly due to the G1, G4, and G6 lensing galaxies. 
To perform better photometry on HLock01, we use {\sc galfit} to model the light distribution of both foreground and background 
components in the SERVS images. 
We use S\'ersic profiles centered at the positions of the detected $HST$/F110W counterparts, and a nearby star was chosen as a PSF model. 
We then measure the flux density of our best-fit model of the lensed images A and C, and use the lens model 
to add the expected light from the other lensed images to obtain the total observed flux.
Modeling the light distribution using {\sc galfit} to separate the fluxes from the SMG and the LBG does not help and will introduce 
significant uncertainties in their measurements, since the spatial separation of the SMG and the LBG in the lensed images A and C is substantially 
lower ($\simeq 0.9^{\prime \prime}$) than the intrinsic PSF in IRAC data ($\simeq 2^{\prime \prime}$ FWHM). 
Finally, in the 8.0~$\mu$m IRAC band the foreground light contamination appears to be much lower than in the other IRAC bands, thus 
we use the $3.8^{\prime \prime}$ aperture photometry for the lensed images A and C provided in the SEIP catalog, 
with the appropriate aperture corrections. 
Again we use the lens model to add the expected light of the lensed images B and D.

\section{Lens modeling}\label{lens}

We use the $\simeq 0.2^{\prime \prime}$ FWHM $HST$/F110W image data to update the lens model already described in \cite{gavazzi2011}. 
The procedure is identical and uses the dedicated code {\sc sl\_fit} \citep[for more details see also:][]{gavazzi2007, 
gavazzi2008, gavazzi2011, gavazzi2012}.
We fit model parameters of simple analytical lensing potentials and model background galaxies as simple 
elliptical S\'ersic profiles. The lensing potential is primarily constrained by the $HST$ data, with the highest resolution 
and S/N. The mass distribution is then held fixed in order to fit for the parameters 
defining the light distribution in the other channels. 
We assume the deflector to be made of an isothermal elliptical mass distribution centered on the galaxy G1 and we also 
include the perturbing galaxies G2, G3 and G4 as point masses centered on the substructure light emission. 
We allow for the presence of a core radius that softens the inner mass distribution in each case. Unlike in \citet{gavazzi2011} we do 
not place masses at G5 and G6, since they have a negligible impact on the mass model and their masses are essentially unconstrained. 
G1, being by far the most massive galaxy in the vicinity, is assumed to be at the center of the group-scale total mass distribution.
The collective effect of a few possible perturbing galaxies 10-20$\arcsec$ South of G1 may induce some external shear, which will 
contribute to the quadrupole of the mass distribution, but having too few constraints spanning too small a radial range around G1, 
we assume that the ellipticity of the mass distribution centered on G1 will absorb the total quadrupole. 
Higher order effects (like $m=1$ or $m=3$ multipoles) would also be hard to constrain with the current data. 

The lensed features exhibiting a cross-like (or barely fold-like) configuration requires a source relatively close to the optical axis, 
and thus, relatively far for a widely opened main astroid caustic. We do not expect much magnification nor huge spatial variations 
of the magnification over the extent of the source. This contrasts with a cusp configuration \citep[see e.g. the cluster 
MS~0451.6-0305 in][]{mackenzie2014}.

\begin{figure*}[ht]
\centering
\includegraphics[width=135mm,scale=1]{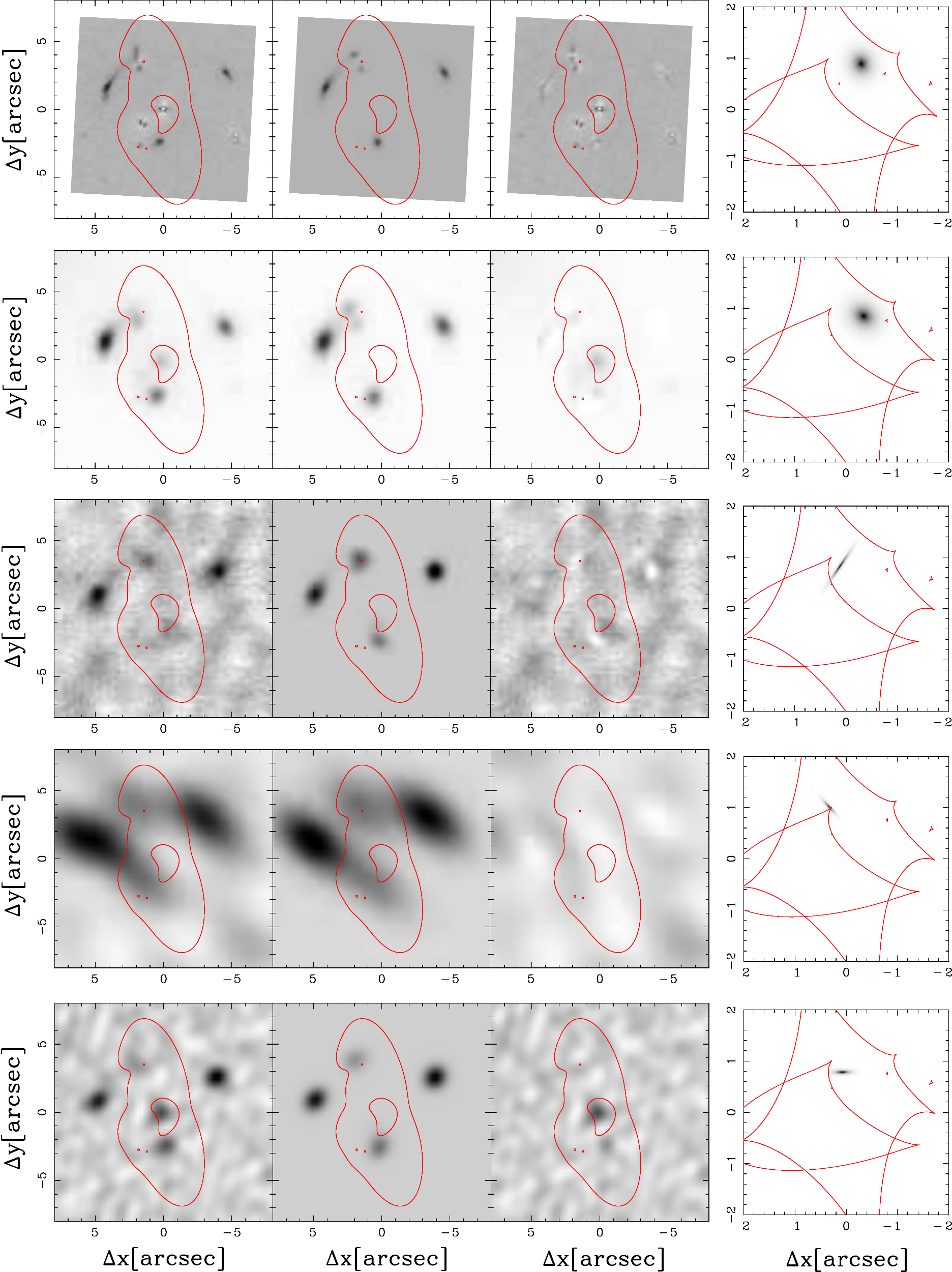}
\caption{Lens inversion results in several bands. Top to bottom: $HST$ F110W; GTC $g$ band; 880~$\mu$m SMA; PdBI CO$\,(J = 5 
\rightarrow 4$); and VLA $1.4\, \rm GHz$ observations. From left to right: input image (foreground deflectors are preliminarily 
subtracted off for $HST$); reconstructed image plane model; residual of input $-$ reconstructed image; and finally the source-plane 
reconstruction. 
All the images are centered on the lensing galaxy G1 ($\alpha = +10$:$57$:$50.959$, $\delta =+57$:$30$:$25.66$, J2000) and 
oriented such that north is up and east is to the left. At long wavelengths, the limited spatial resolution induces large fluctuations 
in the shown reconstructed maximum a posteriori sources (right panels), which can get very elongated (see Figure~\ref{fig:offset} 
for a comparison of more realistic posterior mean sources).
\label{fig:modHST} }
\end{figure*}

Modeling the $HST$ F110W data first, we assume the background source is made of one single elliptical exponential profile for 
which we adjust source plane position, ellipticity, orientation, effective radius, and flux. Before fitting the 
lensed light emission, we performed a fit to the foreground light emission (i.e., G1 to G5) in order to subtract it off. 
The result is shown in the top row panels of Figure \ref{fig:modHST}. 
The formal uncertainty on the recovered Einstein radius (i.e. lens amplitude) is unrealistically small ($\sim 0.2\%$ relative), given 
the high signal-to-noise ratio of the widely extended lensed images. However, large-scale structure mass fluctuations along the line 
of sight, as well as unaccounted for substructures in the lensing mass distribution, should place a lower limit of order 1-2\% on the 
accuracy to which the Einstein radius can be measured. By artificially increasing the pixel rms errors in the F110W imaging data 
by a factor of 10, we are able to mimic this additional source of noise.
As a result, we achieve a one percent accuracy on the recovered Einstein radius as $4\farcs08 
\pm 0\farcs05$, consistent with the previous model of \cite{gavazzi2011}. The core radius is found to be $1\farcs1 \pm 0\farcs1 $, and 
the mass distribution is very elongated, with an axis ratio $b/a = 0.38 \pm 0.03$. The mass of perturbing galaxies is poorly 
constrained: $M_{\rm G2} = ( 6.0^{+5.1}_{-4.3} ) \times 10^{10} \msun$,  $M_{\rm G3} \le 4.4\times 10^{10} \msun$, except the case of 
G4 which induces a splitting of one of the multiple images (B1-B2), yielding  $M_{\rm G4} = ( 9.1^{+6.0}_{-2.6} ) \times 10^{10} 
\msun$. We find a total magnification $\mu = 8.5 \pm 0.5 $ in the $HST$ F110W band. The formal errors on magnification do not account 
for the mass-sheet degeneracy (related to a strong assumption about the mass density slope), which is responsible for 
the differences with the magnification we reported in \cite{gavazzi2011}. Channel-to-channel differential magnifications are, on the 
other hand, more robust and do not depend on the assumed mass distribution. From the best-fit mass model inferred from $HST$ F110W 
data, we extract for the position, ellipticity, size, and flux of an exponential disk in the GTC, VLA, PdBI CO$\,(J = 5 \rightarrow 4$), 
and SMA continuum data. 
The output of the modeling in all the channels is shown in Figure \ref{fig:modHST}. From top to bottom, each row represents 
the results of $HST$ F110W, GTC $g$ band, $880\, \mu \rm m$ SMA, PdBI CO$\,(J = 5\rightarrow 4$), and VLA $1.4\, \rm GHz$ observations.
The overall aspect does not change much with respect to the model of \cite{gavazzi2011}. The geometry of the 
source at long wavelengths is poorly determined and the elongated shape of the best-fit sources in the lower panels is not 
significant.

\end{appendix}

\bibliography{adssample_v2}

\end{document}